\documentclass[
reprint,
superscriptaddress,
 amsmath,amssymb,
 aps,
]{revtex4-1}

\usepackage{graphicx}
\usepackage{dcolumn}
\usepackage{bm}
\usepackage[mathlines]{lineno}
\usepackage[colorlinks=true]{hyperref}
\usepackage{mathrsfs}
\usepackage[usenames, dvipsnames]{xcolor}
\usepackage{float}
\usepackage{ulem}
\usepackage[utf8]{inputenc}
\usepackage{csquotes}
\usepackage{empheq}
\usepackage{multirow}

\usepackage{mathrsfs}

\def\be{\begin{equation}}
\def\ee{\end{equation}}
\def\ba{\begin{eqnarray}}
\def\ea{\end{eqnarray}}

\def\nn{\nonumber}
\def\lf{\left}
\def\rt{\right}

\def\lf{\left}\def\rt{\right}\def\q{\theta} \def\w{\omega}\def\r {\rho} \def\t {\tau}    \def\p {\pi} \def\a {\alpha} \def\s {\sigma}  \def\f {\phi}     \def\l {\lambda}    \def\b {\beta}  \def\m {\mu} \def\pd {\partial}\def\p {\pi}   
      \def\S {\Sigma} \def\D {\Delta}    \def\X {\Xi}   \def\.{\cdot}

\newcommand{\bea}{\setlength\arraycolsep{2pt} \begin{eqnarray}}
\newcommand{\eea}{\end{eqnarray}}

\def\={\equiv}
\def\2{I\hspace{-.1em}I}
\def\3{I\hspace{-.1em}I\hspace{-.1em}I}
\def\4{I\hspace{-.1em}V}

\begin{document}

\title{Emissions of photons near the horizons of Kerr-Sen black holes}

\author{Ming Zhang}
\email{mingzhang@jxnu.edu.cn}
\affiliation{Department of Physics, Jiangxi Normal University, Nanchang 330022, China}
\author{Jie Jiang}
\email{Corresponding author, jiejiang@mail.bnu.edu.cn}
\affiliation{Department of Physics, Beijing Normal University, Beijing 100875, China}
\affiliation{Department of Physics, Jiangxi Normal University, Nanchang 330022, China}

\begin{abstract}
We investigate the escape probabilities of the photons near the horizon of the Kerr-Sen black hole. We find that the escape probabilities of the photons are nonzero in the event horizon limit of the extreme Kerr-Sen black hole if the light sources are near the equator. We show that the escape probability of a photon increases with the radial position of the light source. It is also uncovered that the escape probability decreases if the light source with constant radius moves from the equatorial plane to the pole. Besides, we discover that the escape probability of a photon in the Carter frame is always greater than the one in the locally non-rotating frame, except on the pole, where they are equal.
\end{abstract}

\maketitle

\section{Introduction}\label{introduction}
In a recent work \cite{Ogasawara:2019mir}, Kota Ogasawara et.al. studied the escape probability of a photon emitted from a light source that is at rest in a locally nonrotating frame (LNRF) near the event horizon of the Kerr-Newman black hole. Based on the assumption that all photons are emitted isotropically, they found that the escape probability of a photon emitted from an extreme Kerr-Newman black hole is nonzero in the condition that $a>1/2$ with $a$ the spin per unit mass of the black hole. Specifically, it was obtained that the escape probability becomes a maximum value approximated to 30\% if $a=1$. In other conditions, i.e., $0\leqslant a\leqslant 1/2$ and the Kerr-Newman black hole is nonextreme, the escape probability of the photon in the horizon limit is zero. According to the work, if the Carter constant is zero, the result on the escape probability of a photon emitted isotropically but confined in the equatorial plane of a Kerr black hole in Ref. \cite{Ogasawara:2016yfk} can be reproduced.

Recently, the Event Horizon Telescope collaboration released their observation of the shadow of the supermassive black hole candidate in the center of the giant elliptical galaxy M87 \cite{Akiyama:2019cqa,Akiyama:2019brx,Akiyama:2019sww,Akiyama:2019bqs,Akiyama:2019fyp,Akiyama:2019eap}. The emission ring of the black hole was restored via different calibration and imaging schemes and the observed event-horizon-scale image was shown to be consistent with prospects for a Kerr black hole shadow \cite{Akiyama:2019cqa}. The brightness of the transparent emission region which reveals a dark shadow  is an important issue in the observation of the black hole shadow. To reconstruct the brightness distribution of the source, signals recorded individually at each station of the very long baseline interferometry require a common time reference attained by local atomic clocks which are paired with the Global Positioning System to assure synchronization \cite{Akiyama:2019sww}. The reconstructed image is characterized by a bright ring whose diameter is ~40 ${\mu}$as and an azimuthally  asymmetry brightness with  interior brightness depressions \cite{Akiyama:2019bqs}. The peaked distribution radial brightness profile declines gradually toward the center and the contrast of the brightness at the center compared to the rim was used to confirm the size and width of the crescent shadow \cite{Akiyama:2019eap}. The brightness of the black hole shadow is closely related to the escape probability of the photons \cite{Igata:2019pgb}.

In the region near the black hole, especially the ergosphere between the stationary limit and the event horizon of a rotating black hole, there are typical high-energy events by which some of the energy of the black hole can be extracted and taken away. Penrose suggested a process that a particle dropping into the ergosphere splits into two particles, with one swallowed by the black hole and the other goes back to infinity \cite{Penrose:1971uk}. Besides the particle disintegration, there are two-particle collision editions, whose energy extraction efficiencies are substantially increased, see Refs. \cite{Schnittman:2014zsa,Zaslavskii:2015fqy,Maeda:2018hfi,Zhang:2018gpn,Liu:2018myg,Liu:2019wvp} for recent instances. To a certain extent, the observability of these high-energy events depends on the escape probability of the massive or massless particles from the region near the black hole \cite{Ogasawara:2019mir}.

Besides, the first gravitational waves signal from a binary neutron  star (designated GW170817) merger was observed to be accompanied by a short gamma-ray burst (designated GRB 170817A) \cite{GBM:2017lvd}, after the first direct detection of gravitational waves and the first observation of a binary stellar-mass black hole merger by the  Laser Interferometer Gravitational-Wave Observatory (LIGO) and Virgo \cite{Abbott:2016blz}. A bright optical transient (identified as AT 2017gfo), X-ray and radio emission were also discovered across the electromagnetic spectrum \cite{GBM:2017lvd}. This renders that the study of the emission rate, or escape probability, of the electromagnetic waves from a massive compact object like a neutron star or a black hole benefits the multi-messenger observations of the massive merger.

These observation significances motivate us to further study the emission of the photons in the modified gravities, among which the Kerr-Sen solution is of special interest. Comparing to the Kerr-Newman black hole, the Kerr-Sen black hole owns many distinct properties. The Kerr-Sen black hole is algebraically type-A, yet the Kerr-Newman black hole  is algebraically type-D \cite{Griffiths2009gravitation,Burinskii:1995hk}; besides, the Kerr-Sen spacetime owns the additional dual axion pseudoscalar field and  dilaton scalar field \cite{Wu:2020xxx}. Different characteristics of the capturing and scattering of photons between the  Kerr-Sen spacetime background and the Kerr-Newman background were presented in \cite{Hioki:2008zw}.  Other differences on the  evaporation \cite{Koga:1995bs} and  the gyromagnetic ratios \cite{Horne:1992zy} between these two spacetimes were also reported \cite{Siahaan:2015ljs}. Recently, in \cite{Zhang:2020tfz}, we explored the escape probabilities for both massless photons and massive particles being at rest in a locally non-rotating frame (LNRF) from the Kerr-Sen black hole \cite{Sen:1992ua}. The particle source was put on the equatorial plane and we assumed that the emitted particles are confined in the equatorial plane. One of our findings is that the escape probabilities of massless photons and massive particles share qualitatively similar properties.  In this paper, as a generalization of the work, we will consider that the photons escape isotropically from an equatorial source and are not confined in the equatorial plane in the well-known Kerr-Sen black hole background. Different to the results on the escaping properties of photons in the Kerr-Newman background in Ref. \cite{Ogasawara:2019mir}, as to be seen in what follows, ours here show several new ones, including that all the extreme Kerr-Sen black holes own nonzero equatorial photon escape probabilities in the horizon limit. In section \ref{2}, we will define the emission angles of the photons on the equatorial plane of the Kerr-Sen spacetime. In section \ref{3}, we will get the escape cones and evaluate the escape probabilities for the photons on the equatorial plane of the extreme/nonextreme Kerr-Sen black hole background. In section \ref{off} we will study the escape cones and escape probabilities of the photons emitted from a source on the off-equatorial plane. We will also compare the emission of the photons in the LNRF with that in the Carter frame.The last section will be devoted to our remarks.

\section{Emission angles of photons in the Kerr-Sen spacetime}\label{2}
The Kerr-Sen line element in the low-energy effective field theory for heterotic string theory can be written in the Boyer-Lindquist coordinates as \cite{Sen:1992ua}
\be
\begin{aligned}\label{ksm}
ds^{2}=&-\frac{\D\r_{c}^{2}}{\X}dt^{2}+\frac{\r_{c}^{2}}{\D}dr^{2}+\r_{c}^{2}d\q^{2}+\frac{\X\sin\q}{\r_{c}^{2}}(d\f-\w dt)^{2}\\=&-\frac{\Delta-a^2\sin^2\theta}{\Sigma}dt^2+\frac{\Sigma}{ \Delta} dr^2+\Sigma d \theta^2\\&+\frac{\Xi \sin^2(\theta)}{\Sigma}d\phi^2-\frac{4Mra\sin^2 \theta}{\Sigma}dtd\phi,
\end{aligned}
\ee
where
\be
\begin{aligned}
\rho^2=r^2+a^2\cos^2\theta,\quad\quad&\rho_c^2=\rho^2+2cr=\Sigma,\\
\delta=r^2+a^2+2cr,\quad\quad&\Delta=\delta-2Mr,\\
\Xi=\delta^2-\Delta a^2\sin^2\theta,\quad \,&\w=2Mar/\Xi,\\
c=Q^{2}/(2M),\quad\quad\quad\,\,\,&a=J/M,
\end{aligned}\nn
\ee
with $M,\,Q\,,J\,,c$ are mass, $U(1)$ charge, angular momentum, and twist parameter of the black hole, respectively. The Kerr geometry can be recovered once we set $c=0$. $\D=0$ yields the Cauchy horizon and the event horizon of the black hole as
\begin{align}
r_\text{I}&=M-c-\sqrt{\left(M-c\right)^2-a^2},\\ r_\text{H}&=M-c+\sqrt{\left(M-c\right)^2-a^2}.
\end{align}
To ensure the regularity of the horizons and avoid the occurrence of naked singularity, we must keep
\be
0\leqslant a \leqslant a+c\leqslant M.
\ee
The black hole becomes extreme if $a+c=M$.

Using the Hamilton-Jacobi method or solving the geodesic equation, one can obtain the four-momentum of a photon on geodesic orbit expressed in terms of the first-order differential system as
\begin{subequations}
\bea
\S p^t&=&\frac{\delta}{\Delta}(e\delta-al)-a(ae\sin^2\theta-l),\\
\S p^r&=&\sigma_r\sqrt{\bar{\mathcal{R}}(r)},\\
\S p^\theta&=&\sigma_\theta\sqrt{\bar{\Theta}(\theta)},\\
\S p^\phi&=&\frac{a}{\Delta}(e\delta-al)-\csc^2\theta(ae\sin^2\theta-l),
\eea
\end{subequations}
where
\bea
\bar{\mathcal{R}}(r)&=&(E\delta-aL)^2-\Delta\left[\mathcal{Q}+(L-aE)^2\right]\nn,\\
\bar{\Theta}(\theta)&=&\mathcal{Q}-(L^2\csc^2\theta-a^2E^2)
\cos^2\theta\nn
\eea
are the radial effective potential and the latitude angular effective potential, which, after introducing dimensionless parameters
\begin{align}
b=\frac{l}{e}, ~~ q=\frac{{\mathcal{ Q}}}{e^2},
\end{align}
can be rewritten as
\bea
\mathcal{R}(r)&\equiv&\frac{\bar{\mathcal{R}}(r)}{e^2}=(ab-\delta)^2-\Delta\left[q+(b-a)^2\right],\label{rep}\\
\Theta(\theta)&\equiv&\frac{\bar{\Theta}(\theta)}{e^2}=q-(b^2\csc^2\theta-a^2)\cos^2\theta.\label{anp}
\eea
$E\,,L\,,\mathcal{Q}$ are conserved energy, conserved angular momentum parallel to the black hole symmetry axis and Carter constant \cite{Carter:1968rr} corresponding to the Killing vectors $\partial_{t}$, $\partial_{\f}$ and the Killing-Yano tensor field, respectively. $\s_{r},\,\s_{\q}=\pm 1$ determine the direction of the photon trajectory and they are independent of each other. The four-momentum of the photon is defined by $p^{a}=(\pd/\pd\l)^{a}$ with $\l$ the affine parameter related with the proper time $\t$ of the photon with rest mass $\mu$ by  $\l=\t/\m$.

Rather than describing physics in the inconvenient  Boyer-Lindquist coordinates for the Kerr-Sen metric \eqref{ksm}, we would like to introduce the orthonormal tetrad
\begin{subequations}
\bea
e^{\,\,(t)}_\mu&=&\left(\sqrt{\frac{\Delta  \Sigma }{\Xi }},0,0,0\right),\\
e^{\,\,(r)}_\mu&=&\left(0,\sqrt{\frac{\Sigma }{\Delta }},0,0\right),\\
e^{\,\,(\theta)}_\mu&=&(0,0,\sqrt{\Sigma },0),\\
e^{\,\,(\varphi)}_\mu&=&\left(-\frac{2 a M r \sin \theta }{\sqrt{\Xi  \Sigma }},0,0,\sin \theta \sqrt{\frac{\Xi }{\Sigma }}\right),
\eea
\end{subequations}
carried by an observer rotating with the black hole, i.e., an LNRF \cite{Bardeen:1972fi}, so that we can use relations
\begin{align}
e^{(a)}&=e_{\m}^{\,\,(a)}dx^{\m},\\
e_{(a)}&=e^{\m}_{\,\,(a)}\frac{\pd}{\pd x^{\m}}
\end{align}
to transform the basis vectors back and forth between Boyer-Lindquist coordinate frame and the LNRF frame.

\begin{figure}[!htbp] 
   \centering
   \includegraphics[width=2.8in]{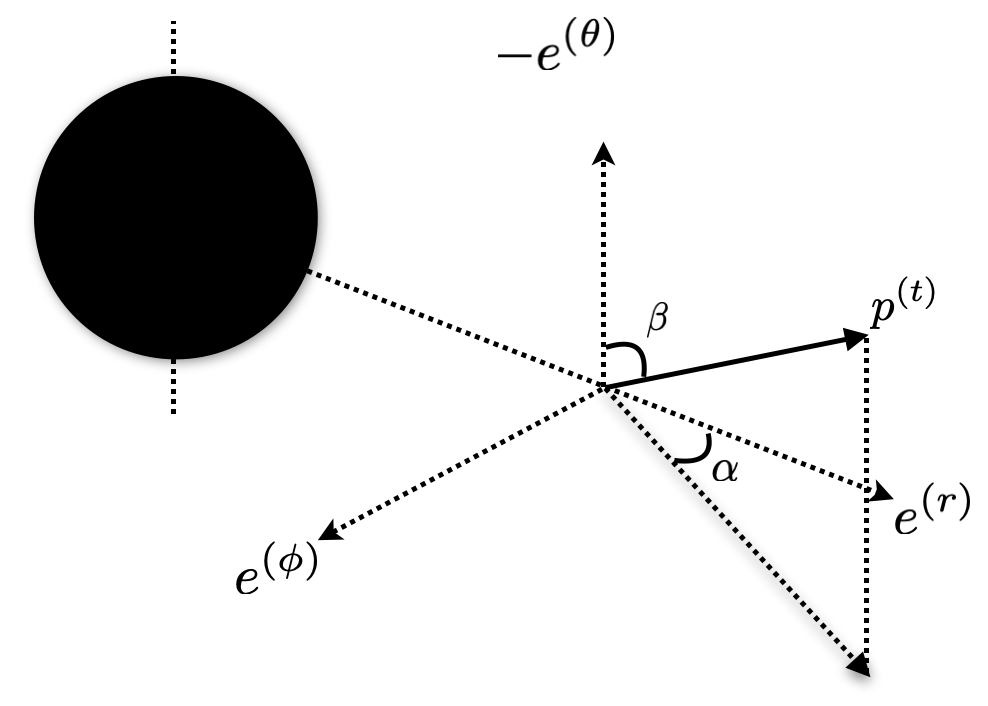}
   \caption{Projection of the momentum $p^{(t)}$ in the observer's LNRF for a photon. We choose $e^{(\theta)}$ as the southward direction. $\a$ is positive in the direction $e^{(\f)}$.}
   \label{schematic0}
\end{figure}

The physically measured  LNRF components of the photon's four-momentum can be got by 
\be
p^{(a)}=p^{\m}e_{\m}^{\,\,(a)},
\ee
which gives
\begin{subequations}
 \begin{align}
p^{(t)}=&\sqrt{\frac{\Delta }{\Xi  \Sigma }} \left[\frac{\delta  (\delta  e-a l)}{\Delta }+a \left(l-a e \sin ^2\theta \right)\right],\label{yy1}\\
p^{(r)}=&\sigma _r \sqrt{\frac{\mathcal{R}}{\Delta  \Sigma }},\label{yy2}\\
p^{(\theta)}=&\sigma _s\sqrt{\frac{\Theta }{\Sigma }} ,\label{ptheta2}\\
p^{(\phi)}=&\sin\q \left[\frac{2 a M r \left(a^2 \Delta  e \sin ^2\theta +a l (\delta -\Delta )-\delta ^2 e\right)}{\Delta  \Sigma  \sqrt{\Xi  \Sigma }}\right.\\&\left.+\frac{\left(\Delta  l \csc ^2\theta-a (a l+e (\Delta -\delta ))\right)}{\Delta  \Sigma\sqrt{\Sigma/\Xi } }\right].
\end{align}
\end{subequations}

Considering a light source staying at rest with a radial position $r=r_{*}$ on the equatorial plane of the Kerr-Sen black hole, the emission angles $(\a,\b)$ of a photon concerning the LNRF can be defined by
\begin{subequations}
 \begin{align}
p^{(r)}&=p^{(t)}\cos\a\sin\b,\\
p^{(\q)}&=-p^{(t)}\cos\b,\\
p^{(\f)}&=p^{(t)}\sin\a\sin\b,
\end{align}
\end{subequations}
where we have used the relation $p^{(a)}p_{(a)}=0$ for the massless photon and we have $p^{(t)}>0$ due to the forward-in-time condition. As shown in Fig. \ref{schematic0}, the angular coordinate $\a$ is the angle between $e^{(r)}$ and the projection of $p^{(t)}$ on the $e^{(r)}-e^{\f}$ plane, the angular coordinate $\b$ is the angle between $-e^{\q}$ and $p^{(t)}$. The domains of $\a$, $\b$ are $\lf[-\p,\,\p\rt]$ and $\lf[0,\,\pi\rt]$, respectively. Starting from this definition, the emission angles can be solved as
\begin{subequations}
 \begin{align}
\sin\a&=\frac{p^{(\f)}}{\sqrt{\left(p^{(r)}\right)^2+\left(p^{(\f)}\right)^2}}\label{ans1},\\
\tan\a&=\frac{p^{(\f)}}{p^{(r)}}\label{ans2},
\end{align}
\end{subequations}
and
\begin{subequations}
 \begin{align}
\sin\b&=\frac{\sqrt{\left(p^{(r)}\right)^2+\left(p^{(\f)}\right)^2}}{p^{(t)}},\label{ans3} \\
\cos\b&=-\frac{p^{(\q)}}{p^{(t)}}\label{ans4},
\end{align}
\end{subequations}
which show that the emission angles are associated with some key parameters as
\begin{align}
\a=&\a(\s_r,\,b,\,q,\, r_*),\\
\b=&\b(\s_\theta,\,b,\,q,\,r_*).
\end{align}
The range of the emission angles can be appointed as
\begin{subequations}
\begin{align}
-\pi&\leqslant\a(\s_r=-1,\,b\leqslant 0,\,q,\, r_*)\leqslant -\frac{\pi}{2},\\
\frac{\pi}{2}&\leqslant\a(\s_r=-1,\,b\geqslant 0,\,q,\, r_*)\leqslant \pi,\\
-\frac{\pi}{2}&\leqslant\a(\s_r=1,\,b\leqslant 0,\,q,\, r_*)\leqslant 0,\\
0&\leqslant\a(\s_r=1,\,b\geqslant 0,\,q,\, r_*)\leqslant \frac{\pi}{2},
\end{align}
\end{subequations}
and
\begin{subequations}
\begin{align}
0&\leqslant\b(\s_\q=-1,\,b,\,q,\, r_*)\leqslant \frac{\pi}{2},\\
\frac{\pi}{2}&\leqslant\b(\s_\q=1,\,b,\,q,\, r_*)\leqslant \pi.
\end{align}
\end{subequations}
Note that $\a=\pi$ is equivalent to $\a=-\p$ and both of them correspond to $b=0$. The photon moves on the equatorial plane for $q=0$.

\section{Escape cones and escape probabilities of photons escaping to infinity}\label{3}
To facilitate our readers, we here will use most of our denotations as those in Ref. \cite{Ogasawara:2019mir}. We set $M=1$. As the light source we choose is located on the equatorial plane of the Kerr-Sen black hole, we have $q\geqslant 0$ according to Eq. \eqref{anp}. By solving $\mathcal{R}=0$, the radial turning points of the photon can be obtained as
\begin{align}
b&=b_1(r)\\&=\frac{\sqrt{a^4 \Delta -2 a^2 \delta  \Delta +a^2 \Delta  q+\delta ^2 \Delta -\Delta ^2 q}-2 a  r}{\Delta -a^2},
\end{align}
and
\begin{align}
b&=b_2(r)\\&=\frac{-\sqrt{a^4 \Delta -2 a^2 \delta  \Delta +a^2 \Delta  q+\delta ^2 \Delta -\Delta ^2 q}-2 a  r}{\Delta -a^2}.
\end{align}
$b_{2}$ diverges at $r_{b_{2}}=\sqrt{c^2-2 c-a^2+1}-c+1$. To make $R\geqslant 0$, we should have
\be
b\leqslant b_1\quad \mathrm{for}\quad r_H\leqslant r<r_{b_{2}},
\ee
\be
b_2\leqslant b\leqslant b_1 \quad \mathrm{for}\quad r> r_{b_{2}}.
\ee

\begin{figure}[!htbp] 
   \centering
   \includegraphics[width=2.8in]{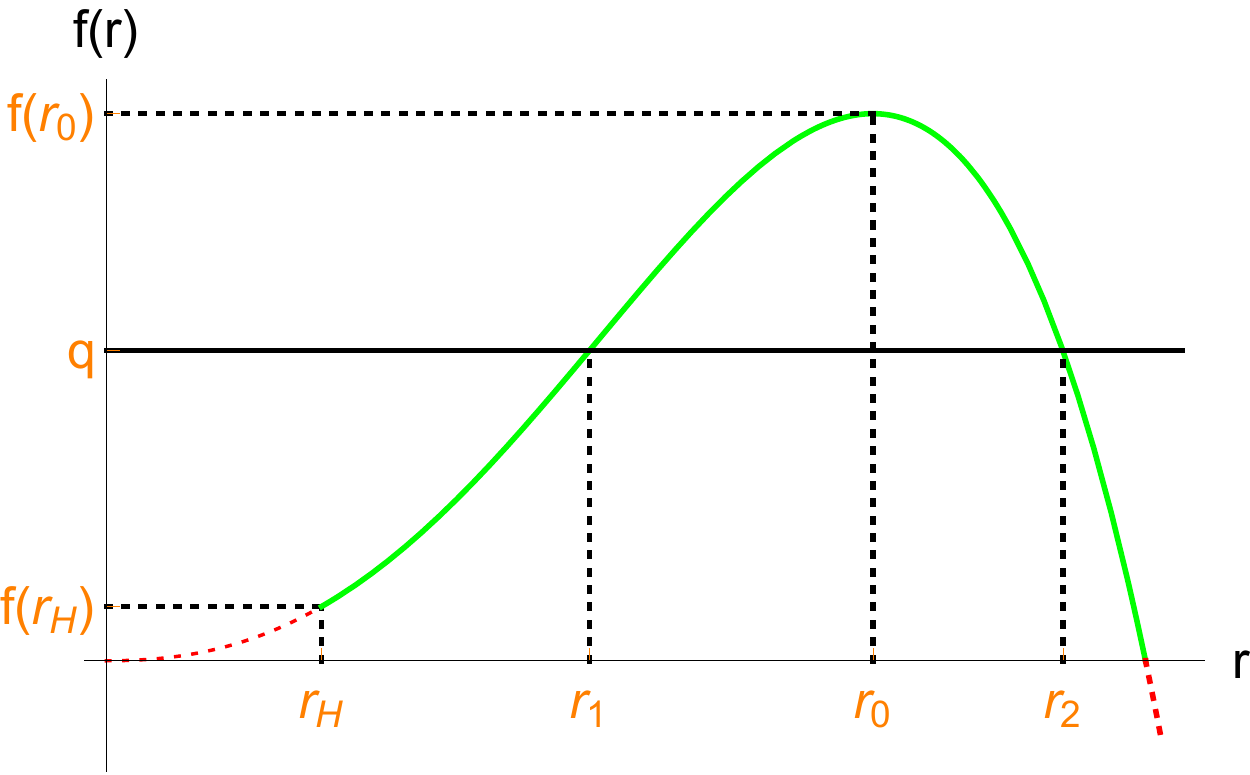}\\
   \includegraphics[width=2.8in]{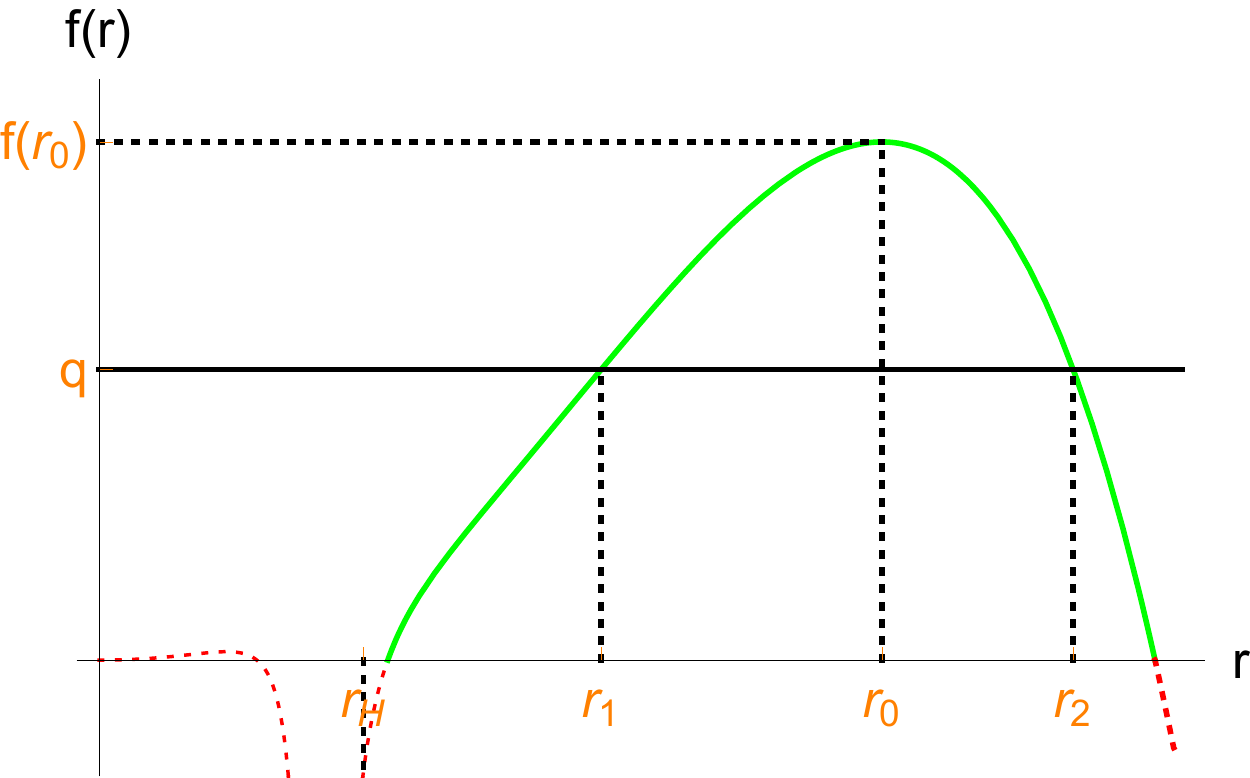}
   \caption{Schematic diagrams of $f(r)$. $r_{1}$ and $r_{2}$ are shown to be roots of Eq. \eqref{carsol}. Top: $f(r)$ for extreme Kerr-Sen black hole. Bottom: $f(r)$ for nonextreme Kerr-Sen black hole.}
   \label{schematic2}
\end{figure}

The extreme values of $b_{i}\,(i=1,2)$ are 
\be
\begin{aligned}
&b^{\textrm{s}}_i \equiv b_i(r_i)\\&=-\frac{a^2 (c+r_{i}+1)+r_{i} \left(2 c^2+c (3 r_{i}-2)+(r_{i}-3) r_{i}\right)}{a (c+r_{i}-1)},
\end{aligned}
\ee
where $r_{i}$ is the solution of $b_i'(r_{i})=0$ and $r_{1}<r_{2}$.

 We have the conditions that $q\geqslant 0$ as well as $r_{*}>r_{H}$, but there still exist two possibilities, i.e., $r_{*}\leqslant r_{1}$ and $r_{*}> r_{1}$. To further confirm the relation between $r_{1}$ and $r_{H}$, we need to analyze the equation 
 \be
\begin{aligned}\label{carsol}
q&=f(r)\\&=-\frac{r^2 \left(\left(2 c^2+c (3 r-2)+(r-3) r\right)^2-4 a^2 (c+r)\right)}{a^2 (c+r-1)^2},
\end{aligned}
\ee
which is solved from $b_i'(r)=0$. By calculation, we know that $f(r)$ becomes maximum at $r=r_{0}$ with
\be
0\leqslant r_0 =\frac{1}{2} \left(\sqrt{c^2-10 c+9}-3 c+3\right)\leqslant 3,
\ee
\be
\begin{aligned}
&4\leqslant f(r_0)\\&=\frac{2 \left(\sqrt{c^2-10 c+9}-3 c+3\right)^2 \left(\sqrt{c^2-10 c+9}-b+3\right)}{\left(\sqrt{c^2-10 c+9}-c+1\right)^2}\\&\leqslant 27,
\end{aligned}
\ee
as well as
\be
f(r_{H})=4 a-a^2>0\quad \mathrm{for}\quad a+c=1,
\ee
\be
\begin{aligned}
f(r_{H})&=\frac{(-c+\chi +1)^2 \left(a^2-2 \left(c^2+c \chi +\chi +1\right)\right)}{a^2}\\&<\frac{(-c+\chi +1)^2 \left(a^2-2 \left(c^2+1\right)\right)}{a^2}\\&<0\\&\quad\quad\quad\quad\quad\quad\quad\quad\quad\quad\quad\quad\quad\quad\mathrm{for} \quad a+c<1,
\end{aligned}
\ee
where $\chi=\sqrt{(b-1)^2-a^2}$. We show schematic diagrams for the two kinds of $f(r)$ in Fig. \ref{schematic2}. Different to the Kerr-Newman case in Ref. \cite{Ogasawara:2019mir},  we have $q>0$ at the event horizon for extreme Kerr-Sen black hole and $q<0$ at the event horizon for nonextreme Kerr-Sen black hole. As a result, we can know that there exist three cases about the relations of $r_{1},\,r_{H}\,$ and $r_{*}$, which are
\begin{subequations}
\begin{empheq}{alignat=2}
 &\textrm{Case~(1):}& ~ &r_1< r_{\mathrm{H}}<r_*, \\[1mm]
 &\textrm{Case~(2):}& ~ &r_{\mathrm{H}}\leq r_1<r_*, \\[1mm]
 &\textrm{Case~(3):}& ~ &r_{\mathrm{H}}<r_*\leq r_1.
\end{empheq}
\end{subequations}

\begin{figure*}[!htbp] 
   \centering
   \includegraphics[width=2.3in]{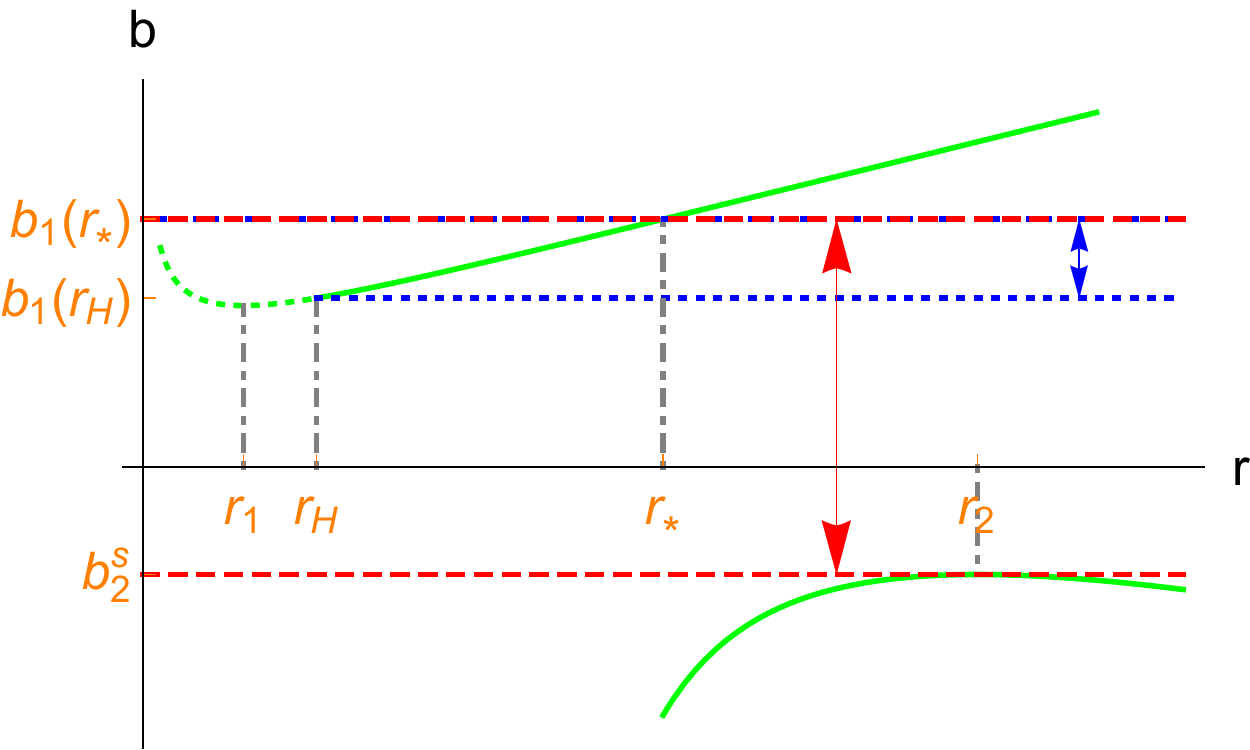}
    \includegraphics[width=2.3in]{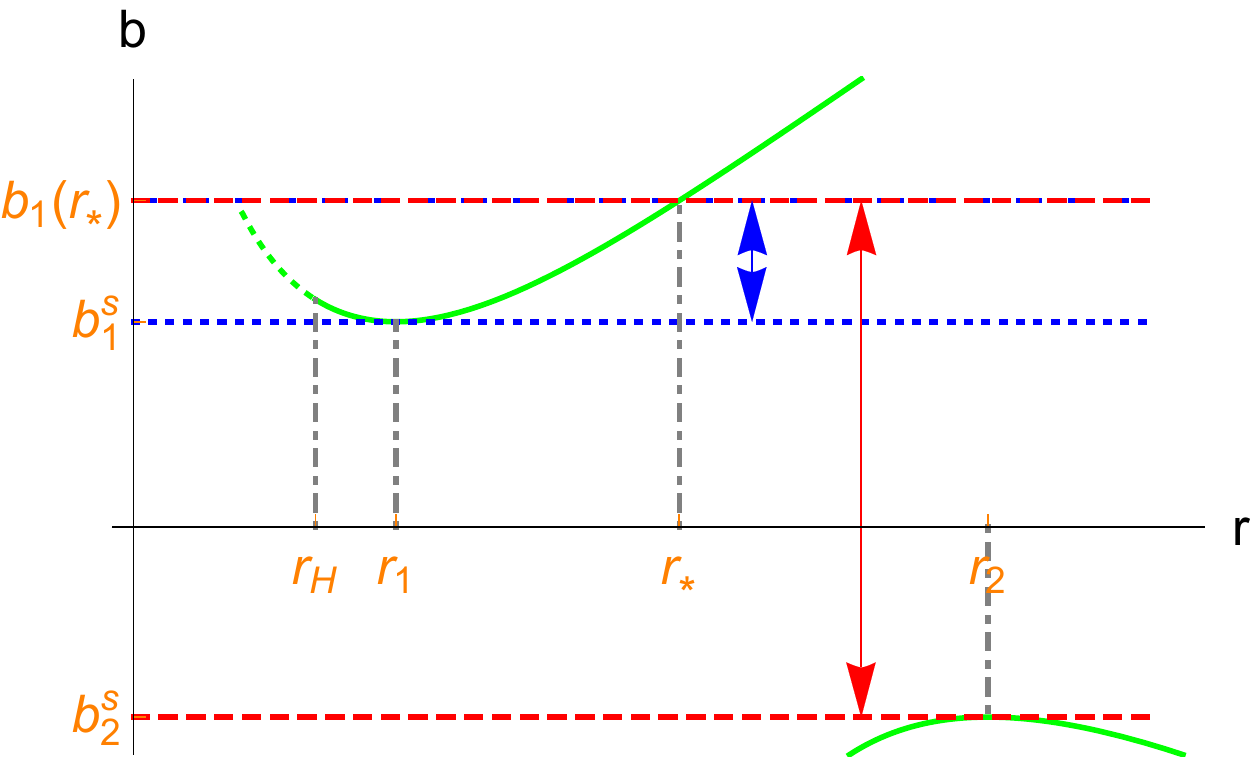}
     \includegraphics[width=2.3in]{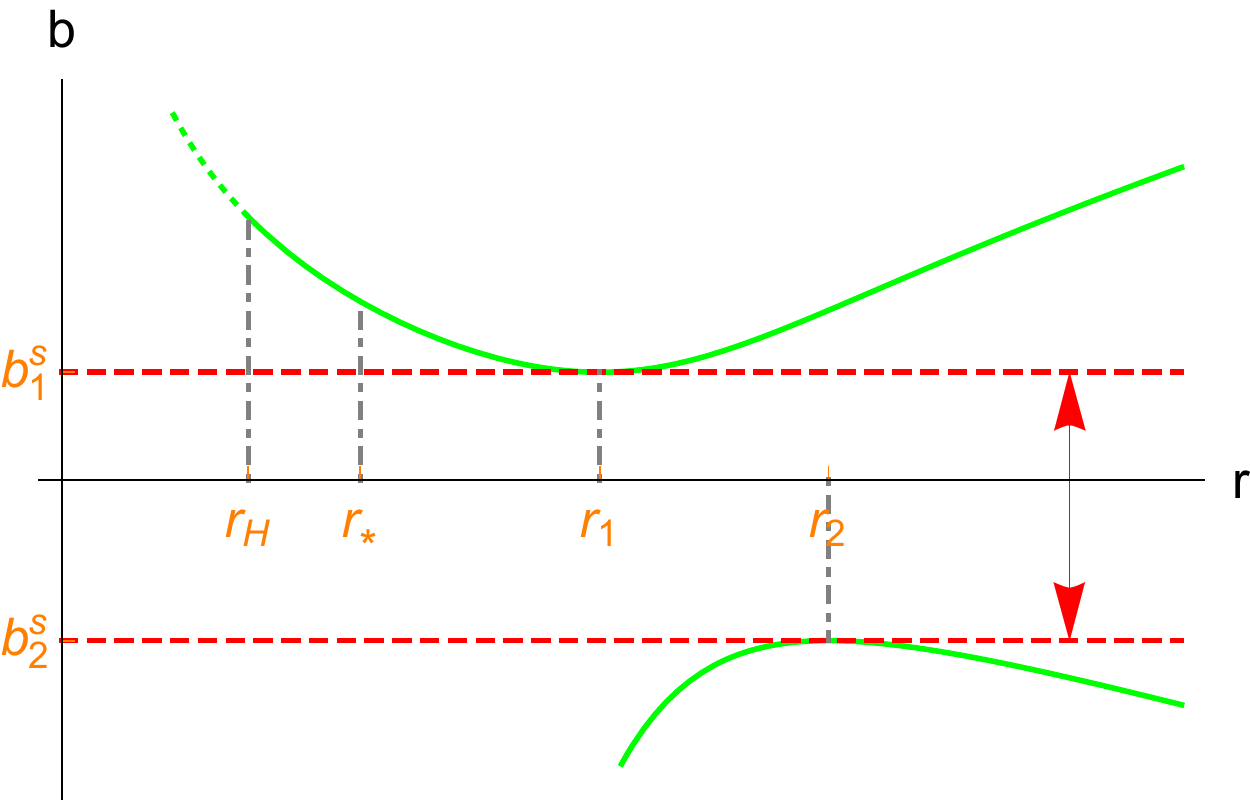}
   \caption{The impact parameters $b_{i}$ are shown for cases (1)-(3) in sequence. The photon with  outward radial momentum can escape from the Kerr-Sen black hole in the range marked by the vertical red arrow. The photon with inward radial momentum can escape from the Kerr-Sen black hole in the range marked by the vertical blue arrow.}
   \label{schematic3}
\end{figure*}

We now determine the boundary of escape cone for a photon from the extreme or nonextreme Kerr-Sen black hole case by case. We will restrict the light source in the range $r_{\mathrm{H}} < r_* \leq r_0$ as the calculation is similar for $r>r_{0}$.

\begin{figure*}[!htbp] 
   \centering
   \includegraphics[width=2.3in]{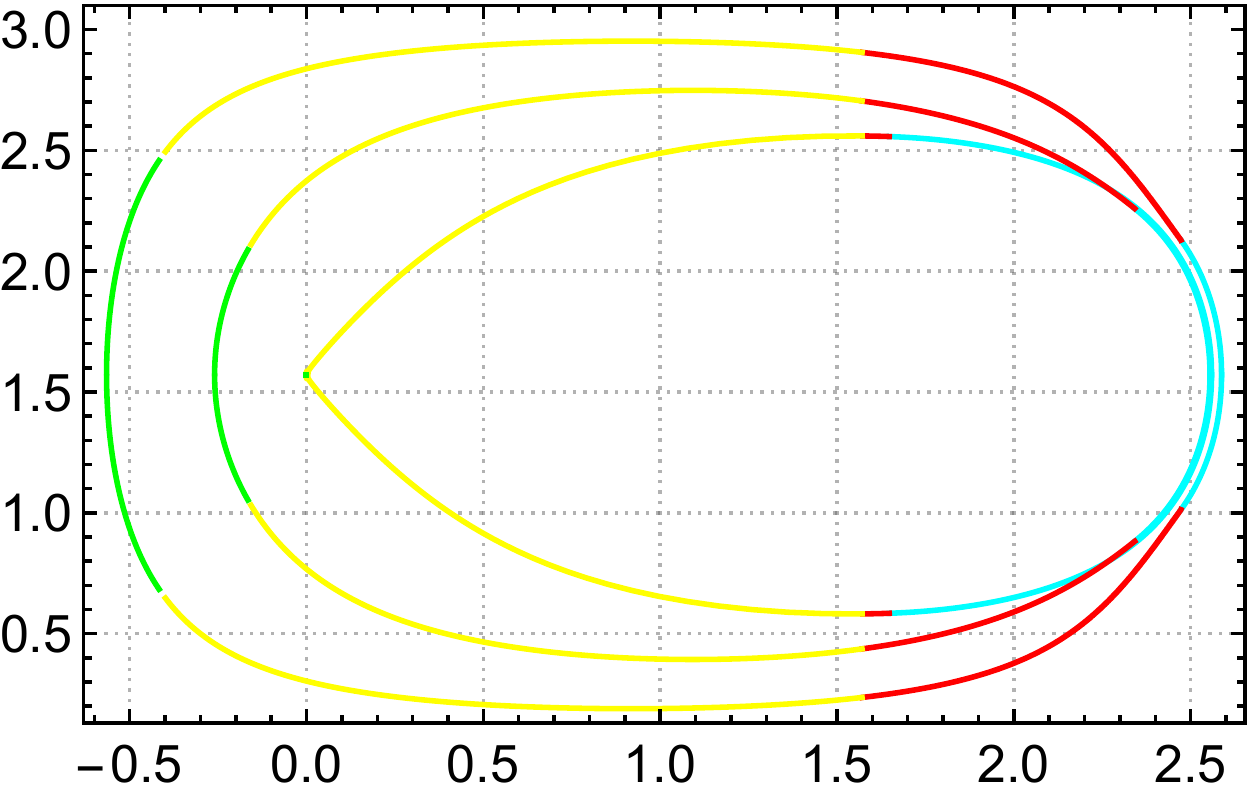}
     \includegraphics[width=2.3in]{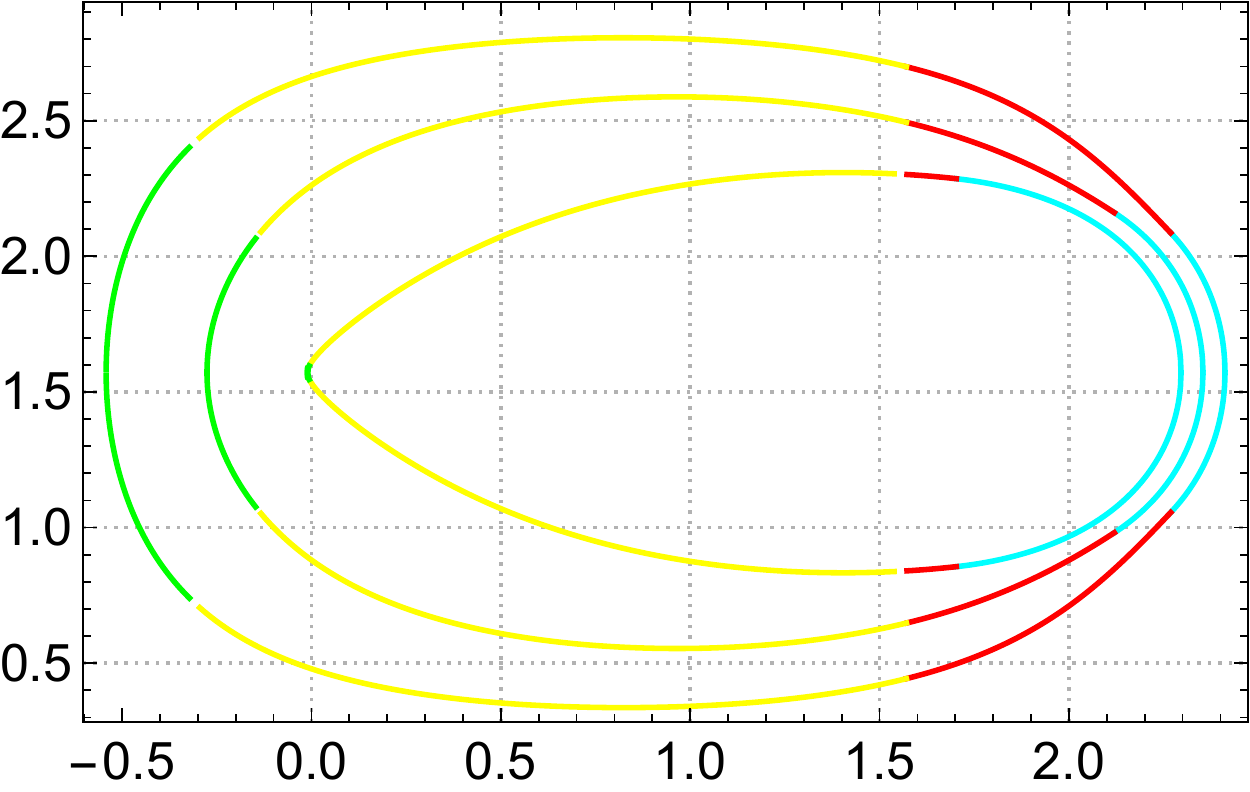}
       \includegraphics[width=2.3in]{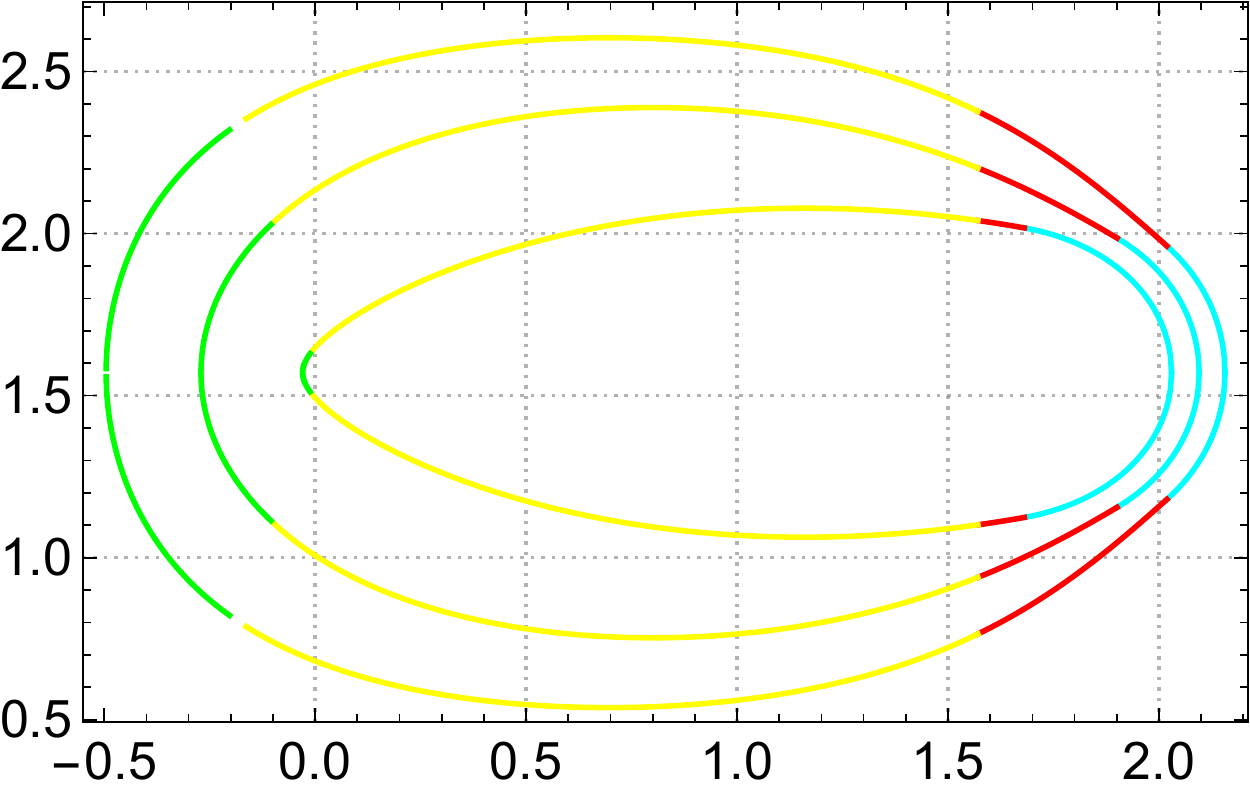}
   \caption{Critical angles for the extreme Kerr-Sen black hole. The vertical axes and horizontal axes stand for $\b_{(i)}$ and $\a_{(i)}$, respectively. Left: $a=0.9, r_*=0.9018, 1.35, 1.8$ from inside to outside. Middle: $a=0.5, r=0.51, 0.75, 1$ from inside to outside. Right: $a=0.2, r=0.21, 0.3, 0.4$. The cyan, red and yellow curves individually stand for the former parts of $\lf(\a_{(i)},\,\b_{(i)}\rt)$ in Eqs. (\ref{eca1}), (\ref{eca2}) and (\ref{ca3}), whose latter parts are represented by the green curves.}
   \label{EscapeConeExtreme}
\end{figure*}

\begin{figure}[!htbp] 
   \centering
   \includegraphics[width=2.8in]{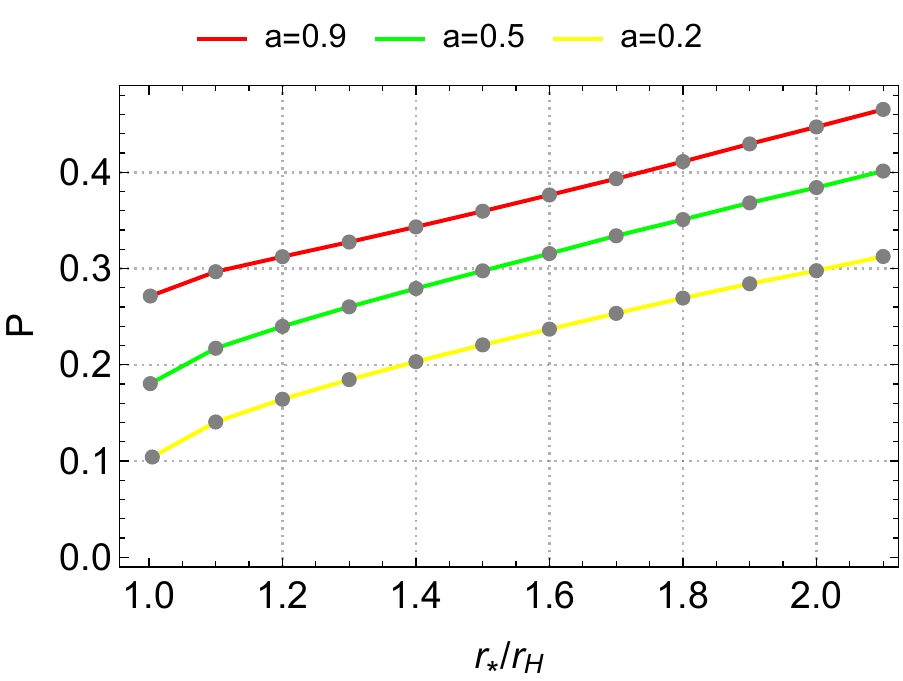}
   \caption{Variations of the escape probabilities with respect to the radial position  for the photons emitted from the extreme Kerr-Sen black hole. }
   \label{epex}
\end{figure}

\begin{figure}[!htbp] 
   \centering
   \includegraphics[width=2.8in]{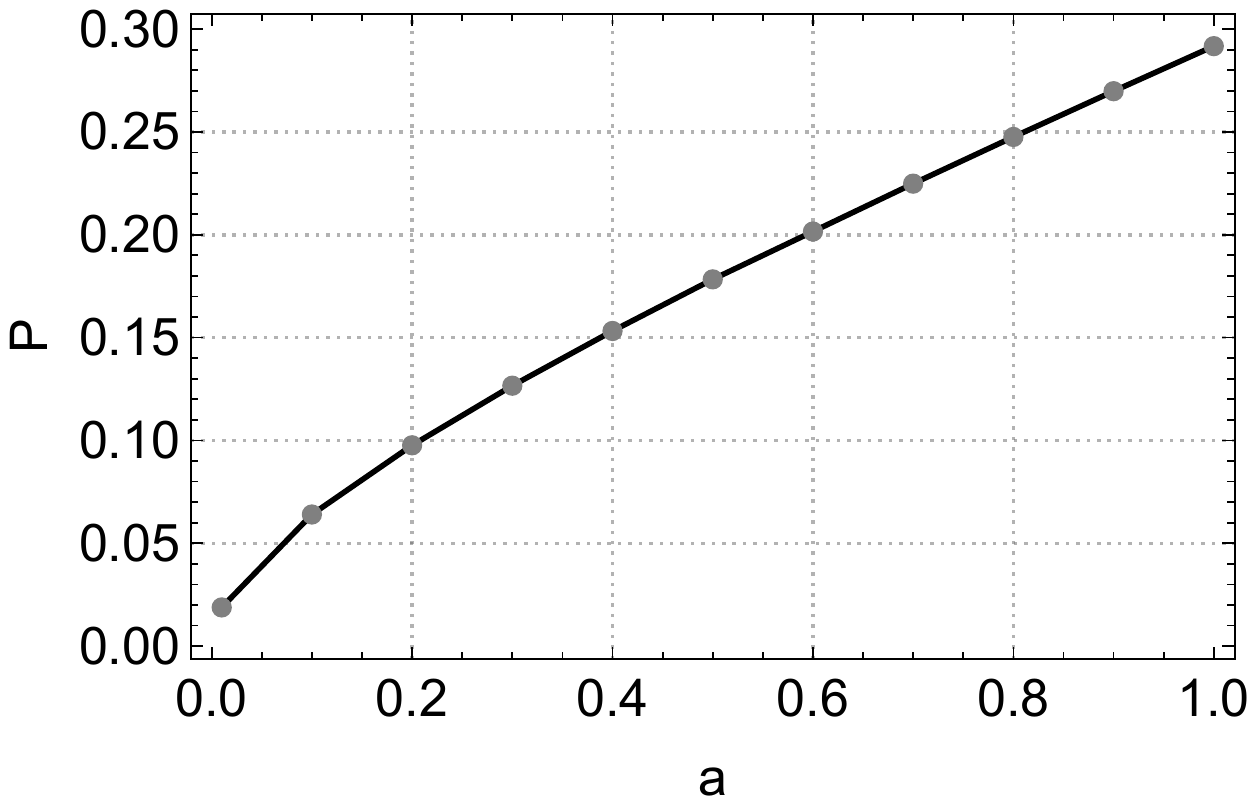}
   \caption{Variations of the escape probabilities with respect to the angular momentum of the extreme Kerr-Sen black hole for the emitted photon in the horizon limit.}
   \label{epexa}
\end{figure}

\begin{figure*}[!htbp] 
   \centering
   \includegraphics[width=2.3in]{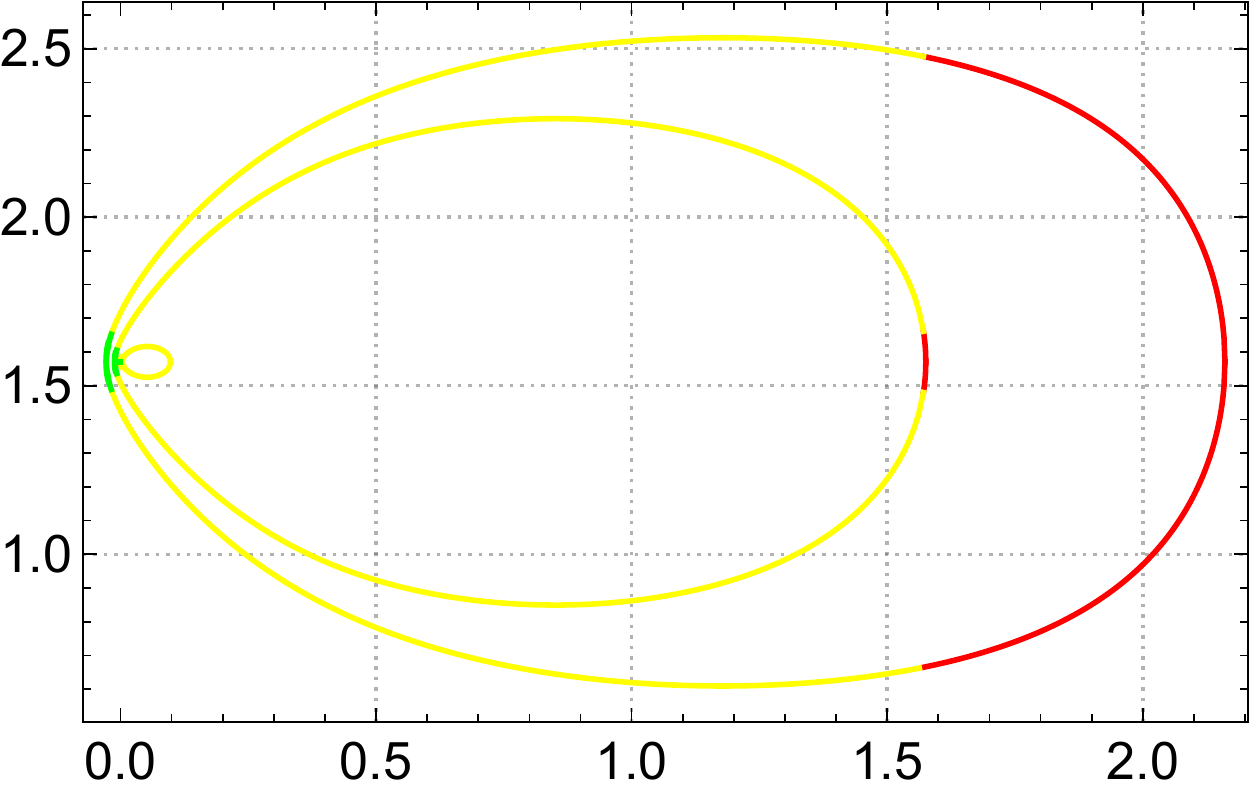}
     \includegraphics[width=2.3in]{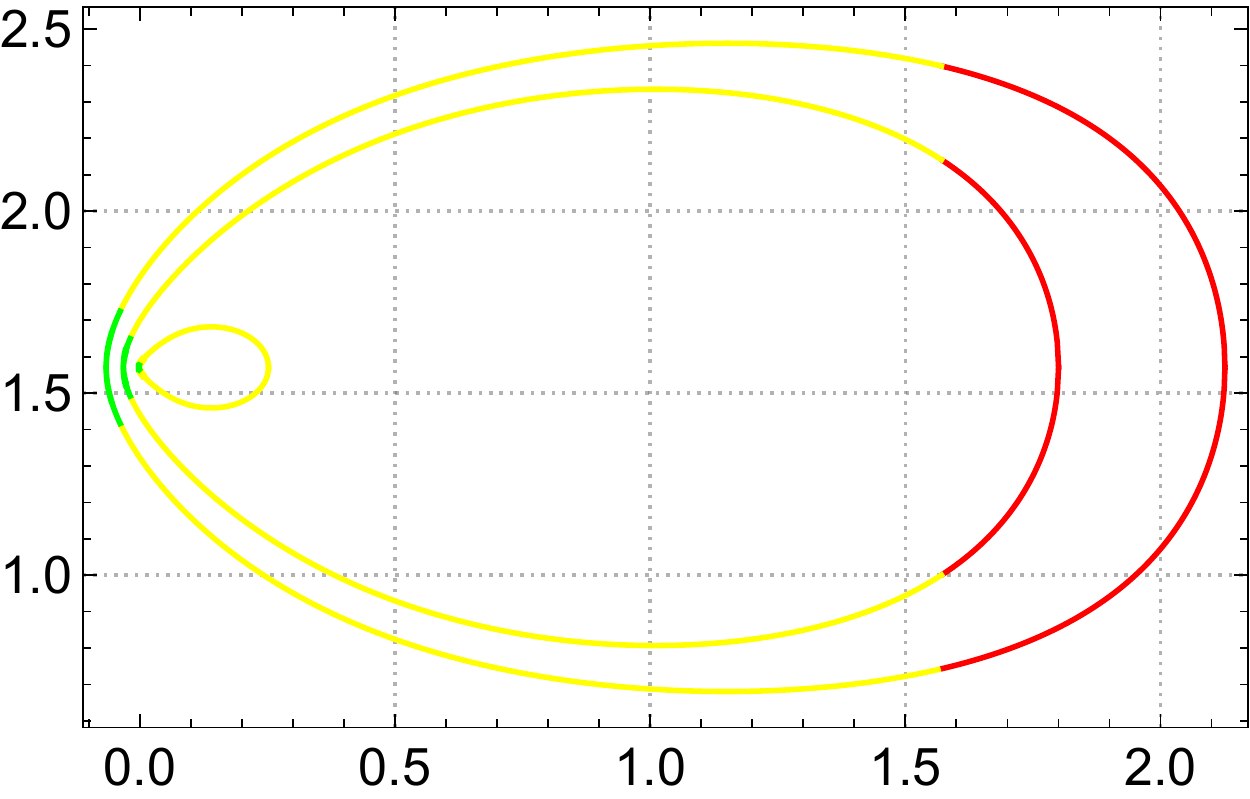}
       \includegraphics[width=2.3in]{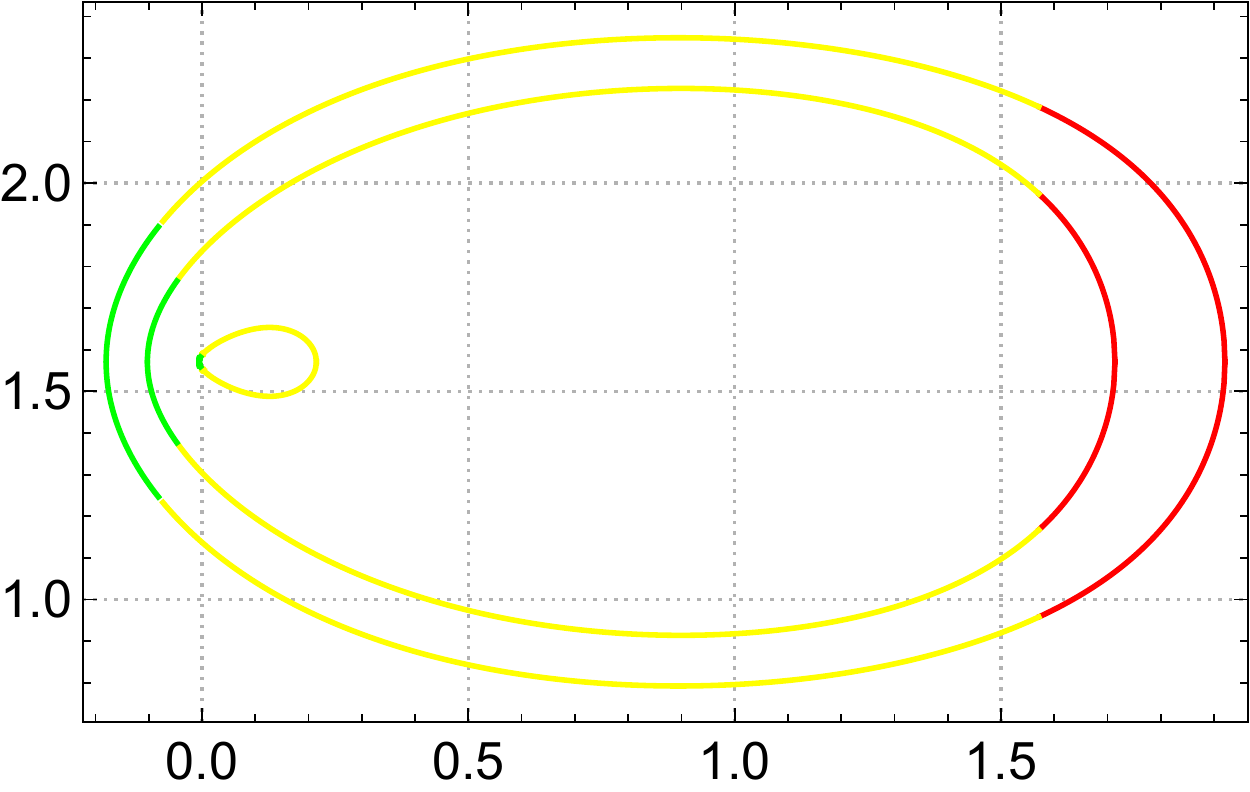}
   \caption{Critical angles for the non-extreme Kerr-Sen black hole with $K=0.999$. The vertical axes and horizontal axes stand for $\b_{(i)}$ and $\a_{(i)}$, respectively. Left: $a=0.98, r_*=1.0253, 1.033, 1.06$ from inside to outside. Middle: $a=0.7, r=0.7386, 0.76, 0.8$ from inside to outside. Right: $a=0.3, r=0.3257, 0.36, 0.4$.  The red and yellow curves individually stand for the former parts of $\lf(\a_{(i)},\,\b_{(i)}\rt)$ in Eqs. (\ref{neca2}) and (\ref{ca3}), whose latter parts are represented by the green curves.}
   \label{nonextcone1}
\end{figure*}

\begin{figure}[!htbp] 
   \centering
   \includegraphics[width=2.8in]{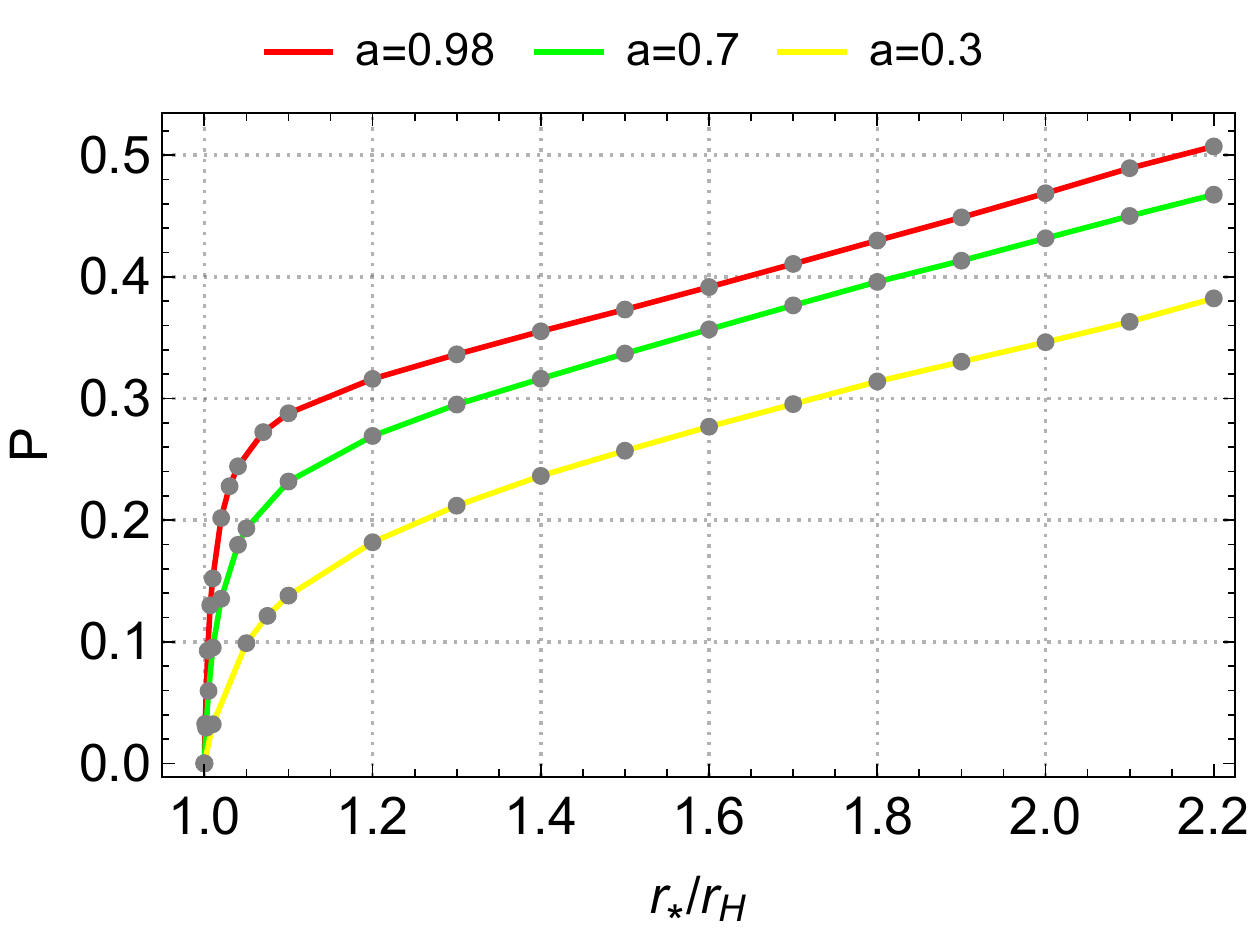}
   \caption{Variations of the escape probabilities with respect to the radial position  for the photons emitted from the nonextreme Kerr-Sen black hole with $K=0.999$.}
   \label{nonextremecharged}
\end{figure}

\begin{figure*}[!htbp] 
   \centering
   \includegraphics[width=2.3in]{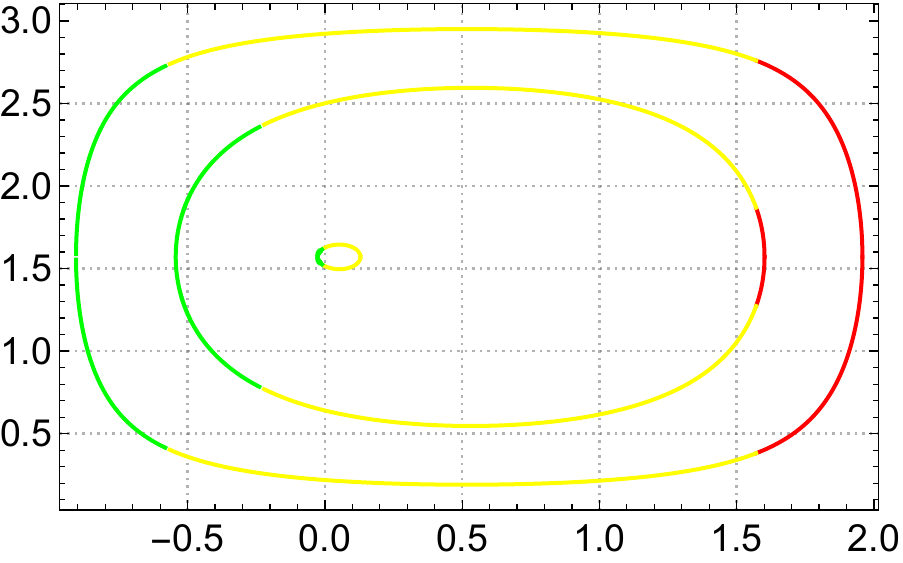}
     \includegraphics[width=2.3in]{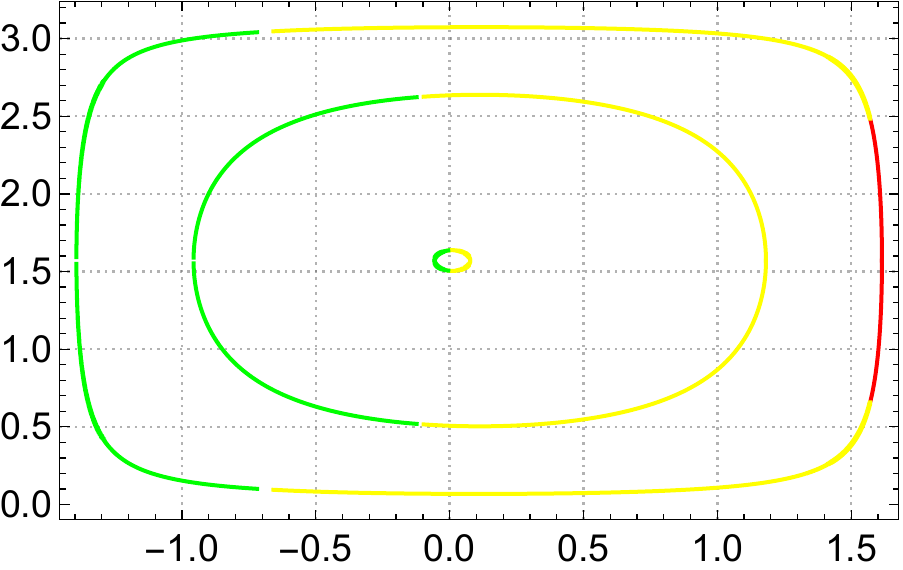}
   \caption{Critical angles for the non-extreme Kerr-Sen black hole with $c=0.5$ and $a=0.4\,(\mathrm{left})\,,a=0.1\,(\mathrm{right})$. The vertical axes and horizontal axes stand for $\b_{(i)}$ and $\a_{(i)}$, respectively. $r_*=1.001 r_H, 1.3r_H, 1.7r_H$ from inside to outside. The red and yellow curves individually stand for the former parts of $\lf(\a_{(i)},\,\b_{(i)}\rt)$ in Eqs. (\ref{neca2}) and (\ref{ca3}), whose latter parts are represented by the green curves.}
   \label{nonextcone2}
\end{figure*}

\begin{figure}[!htbp] 
   \centering
   \includegraphics[width=2.8in]{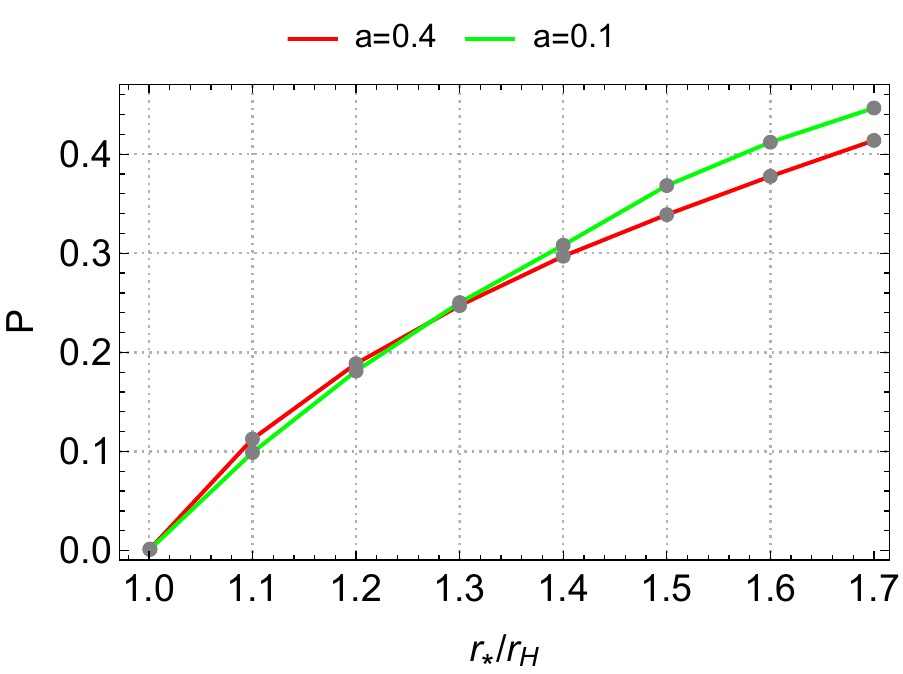}
   \caption{Escape probability of the photon emitted from the nonextreme Kerr-Sen black hole with $c=0.5$.}
   \label{nonextremeuncharged}
\end{figure}

{\it{Case (1)}}
This case appears only for extreme Kerr-Sen black hole as $r_{1}<r_{H}$, which can be observed in the top diagram of Fig. \ref{schematic2}. Combing the left diagram in Fig. \ref{schematic3} with the radial  effective potential \eqref{rep}, we can know that there are two critical conditions that the photons just cannot escape to infinity. One is a radially inward photon with the impact parameter $b=b_{1}(r_{H})$ and the other is a radially outward photon with the impact parameter $b=b_{2}^{s}$. The radial momentum of the photon with the impact parameter $b_{2}^{s}$ vanishes. So the critical angles that a photon can escape from the extreme Kerr-Sen black hole to infinity are
\begin{equation}
\begin{aligned}\label{eca1}
\lf(\a_{(1)},\,\b_{(1)}\rt)_{\mathrm{ext}}=&\lf.\lf(\a,\,\b\rt)\rt|_{\lf[\s_{r}=-1,\,b=b_{1}(r_{H}),\,0\leqslant q<f(r_{H}) \rt]}\\&\cup\lf.\lf(\a,\,\b\rt)\rt|_{\lf[\s_{r}=1,\,b=b_{2}^{s},\,0\leqslant q<f(r_{H}) \rt]},
\end{aligned}
\end{equation}
where $b_{1}(r_{H})=2$.

{\it{Case (2)}}
As $r_{1}>r_{H}$, we have $f(r_{H})\leqslant q<f(r_{*})$ for extreme Kerr-Sen black hole and $0\leqslant q<f(r_{*})$ for nonextreme Kerr-Sen black hole. As shown in the middle diagram of Fig. \ref{schematic3}, photons with impact parameters $b_{1}^{s}$ and $b_{2}^{s}$ can just move inward to a bounded circular orbit with radius $r=r_{1}$ and move outward to the other bounded circular orbit with radius $r=r_{2}$, respectively. As a result, in this case, the critical angles that photons can escape from the extreme Kerr-Sen black hole to infinity are
\begin{equation}
\begin{aligned} \label{eca2}
\lf(\a_{(2)},\,\b_{(2)}\rt)_{\mathrm{ext}}=&\lf.\lf(\a,\,\b\rt)\rt|_{\lf[\s_{r}=-1,\,b=b_{1}^{s},\,f(r_{H})\leqslant q<f(r_*) \rt]}\\&\cup\lf.\lf(\a,\,\b\rt)\rt|_{\lf[\s_{r}=1,\,b=b_{2}^{s},\,f(r_{H})\leqslant q<f(r_*) \rt]},
\end{aligned}
\end{equation}
and the critical angles that photons can escape from the nonextreme Kerr-Sen black hole to infinity are 
\begin{equation}
\begin{aligned}\label{neca2}
\lf(\a_{(2)},\,\b_{(2)}\rt)_{\mathrm{n-ext}}=&\lf.\lf(\a,\,\b\rt)\rt|_{\lf[\s_{r}=-1,\,b=b_{1}^{s},\,0\leqslant q<f(r_*) \rt]}\\&\cup\lf.\lf(\a,\,\b\rt)\rt|_{\lf[\s_{r}=1,\,b=b_{2}^{s},\,0\leqslant q<f(r_*) \rt]}.
\end{aligned}
\end{equation}

{\it{Case (3)}}
There is not any radially inward photon that can escape from the Kerr-Sen black hole to infinity in this case. According to the right diagram in Fig. \ref{schematic3}, the condition that the radially outward photons can go to infinity is $b_{1}^{s}<b<b_{2}^{s}$. Consequently, we get the critical angles of photons escaping from extreme or nonextreme Kerr-Sen black hole to infinity as
\begin{equation}
\begin{aligned}\label{ca3}
\lf(\a_{(3)},\,\b_{(3)}\rt)=&\lf.\lf(\a,\,\b\rt)\rt|_{\lf[\s_{r}=1,\,b=b_{1}^{s},\,f(r_{*})\leqslant q<f(r_0) \rt]}\\&\cup\lf.\lf(\a,\,\b\rt)\rt|_{\lf[\s_{r}=1,\,b=b_{2}^{s},\,f(r_{*})\leqslant q<f(r_0) \rt]}.
\end{aligned}
\end{equation}

According to cases (a)-(c), we obtain the boundary of the escape cone for the photon escaping from the extreme or nonextreme Kerr-Sen black hole to infinity as
\be
\pd S=\bigcup_{i=1,2,3}\lf\{\lf.\lf(\a_{(i)},\b_{(i)}\rt)\rt|_{\s_{\q}=1},\,\lf.\lf(\a_{(i)},\b_{(i)}\rt)\rt|_{\s_{\q}=-1}\rt\}.
\ee
The escape cone $S$ is the solid angles surrounded by the critical angles. Note that the escape cone is symmetric to the equatorial plane of the Kerr-Sen black hole, which means $\b_{(i)}|_{\s_{\q}=-1}=\p-\b_{(i)}|_{\s_{\q}=1}$.

Supposing that photons are emitted isotropically from the light source near the Ker-Sen black hole, the escape probability of the photons can be defined by
\be
P(r_*)\=\frac{1}{4\pi}\int_{S}\mathrm{d}\a \;\mathrm{d}\b\;\sin\b.
\ee
In Ref. \cite{Ogasawara:2019mir}, the escape probability of photons emitted from a light source near the Kerr-Newman black hole was calculated using the coordinates $r_{i}$, as 
\be
P(r_*)\=\sum_{i=1}^{2}\frac{(-1)^{i}}{2\pi}\int_{r_{c,i}}^{r_{0}}\mathrm{d}r_{i}\,\frac{\mathrm{d}\a}{\mathrm{d}r_{i}}\,\cos\b_{i},
\ee
where $r_{c,i}$ are solutions of $f(r)=0$. We realize that the escape probability can be written in form of
\be
P(r_*)\=-\frac{1}{4\pi}\int_{S}\mathrm{d}\a \;\mathrm{d}\cos\b,
\ee
which means the probability is related to the warped escape cone $\a-\cos\b$ whose area can be calculated by  numerical method.

Since the escape cones and  escape probabilities of photons around extreme Kerr-Sen black hole are different from the nonextreme Kerr-Sen black hole case, we will now evaluate them separately.

\subsection{Extreme Kerr-Sen black hole}

The horizon of the extreme Kerr-Sen black hole is $r_{h}=a$. The impact impacter reads
\be
b_{i}^{s}=-\frac{r_{i}^2}{a}-a+2 r_{i}+2,
\ee
where $r_{i}$ can be obtained from 
\be
q=f(r_{i})=-\frac{r_{i}^2 \left(4 a^2-4 a (r_{i}+1)+r_{i}^2\right)}{a^2}.
\ee
We show critical angles for photons emitted from a light source near the extreme Kerr-Sen black hole  in Fig. \ref{EscapeConeExtreme}. We can see that even if the position of the light source approaches the horizon, the escape cone of the photon does not vanish. We further evaluate the escape probabilities of the photons at different positions around extreme Kerr-Sen black hole with different angular momentum in Fig. \ref{epex}. We know that the escape probability of the photon increases with radial distance relative to the horizon. As seen in Fig. \ref{epexa}, the escape probability of the photon in the horizon limit increases with the angular momentum $a$ of the extreme Kerr-Sen black hole. Note that $P\to 0$ for $a\to 0$, and $P\sim 0.292$ for $a=1$ (which is the extreme Kerr black hole case that has been discussed in Ref. \cite{Ogasawara:2019mir}).

\subsection{Nonextreme Kerr-Sen black hole}
We set $c=K-a$ with $0<K<1$ for the nonetreme Kerr-Sen black hole. In Fig. \ref{nonextcone1} and Fig. \ref{nonextremecharged}, we show the escape cones and the escape probabilities of the photons near the horizon of the nonextreme Kerr-Sen black hole with $K=0.999$. We can see that the escape cones shrink and the escape probabilities decrease with the decreasing of the radial distance relative to the horizon of the nonextreme Kerr-Sen black hole. Especially, when the light source approaches to the horizon, the escape cones become vanishing and the escape probabilities become zero. When we set $c=0.5$ for the nonextreme Kerr-Sen black hole, we can see similar results as shown in Figs. \ref{nonextcone2} and \ref{nonextremeuncharged}.

\section{Emission of the photon on the off-equatorial plane}\label{off}
In the former section, we closely followed the Ref. \cite{Ogasawara:2019mir} to calculate the escape cones and escape probabilities of the photons around the extreme Kerr-Sen black hole in the LNRF on the equatorial plane. In what follows, we will further investigate the emission of the photons on the off-equatorial plane. And we will compare the escape of the photons both in the LNRF and in the Carter frame \cite{semerak1996photon,Chang:2020miq}. The Carter frame reads
\begin{subequations}
\bea
e^{\,\,(t)}_\mu&=&\left(\sqrt{\frac{\Delta}{\Sigma}},0,0,-a\sin^{2}\theta\sqrt{\frac{\Delta}{\Sigma}}\right),\\
e^{\,\,(r)}_\mu&=&\left(0,\sqrt{\frac{\Sigma }{\Delta }},0,0\right),\\
e^{\,\,(\theta)}_\mu&=&(0,0,\sqrt{\Sigma },0),\\
e^{\,\,(\varphi)}_\mu&=&\left(-\frac{ a  \sin \theta }{\sqrt{\Sigma }},0,0,\frac{\sin \theta }{\sqrt{\Sigma }}\left(\Sigma+a^{2}\sin^{2}\theta\right)\right).
\eea
\end{subequations}
The components of the photon's four-momentum in the Carter frame can be got as
\begin{subequations}
 \begin{align}
p^{(t)}=&\frac{\left(a^2 \cos 2 \theta -a^2+2 \delta \right) (\delta  e-a l)}{2 \sqrt{\Delta } \Sigma ^{3/2}},\label{xx1}\\
p^{(r)}=&\sigma _r \sqrt{\frac{\mathcal{R}}{\Delta  \Sigma }},\label{xx2}\\
p^{(\theta)}=&\sigma _s\sqrt{\frac{\Theta }{\Sigma }} ,\label{ptheta}\\
p^{(\phi)}=&\frac{1}{\Delta  \Sigma ^{3/2}}\left[\Delta  l \Sigma  \csc \theta-a^2 l (\Sigma -\delta ) \sin \theta\right.\nonumber\\&\left.+a^3 \sin ^3\theta (a l-\delta  e)+a e \sin\theta  \left(\delta ^2-\delta  \Sigma +\Delta  \Sigma \right)\right].
\end{align}
\end{subequations}

The escape cones of the photons in the Carter frame may be similar to those in the LNRF, as the radial effective potentials are the same. However, the escape probabilities of the photons should be different, as the components $p^{(t)}$ and $p^{(\phi)}$ in the Carter frame are different from those in the LNRF, which leads to different emission angles $\alpha$ and $\beta$ for the photons. Setting that the photon source locates at the position $(r_*,\,\theta_*)$, we must ensure that 
\begin{align}
\mathcal{R}(r_*) &\geqslant 0,\\
\Theta(r_*, \theta_*) &\geqslant 0.
\end{align}
These formulas restrict the parameter space to make the emission process physical. Due to the symmetric characteristics of the rotating Kerr-Sen black hole, we have
\begin{equation}
P(r_*,\,\theta_*)=P(r_*,\,\pi-\theta_*).
\end{equation}

\begin{figure}[!htbp] 
   \centering
   \includegraphics[width=2.3in]{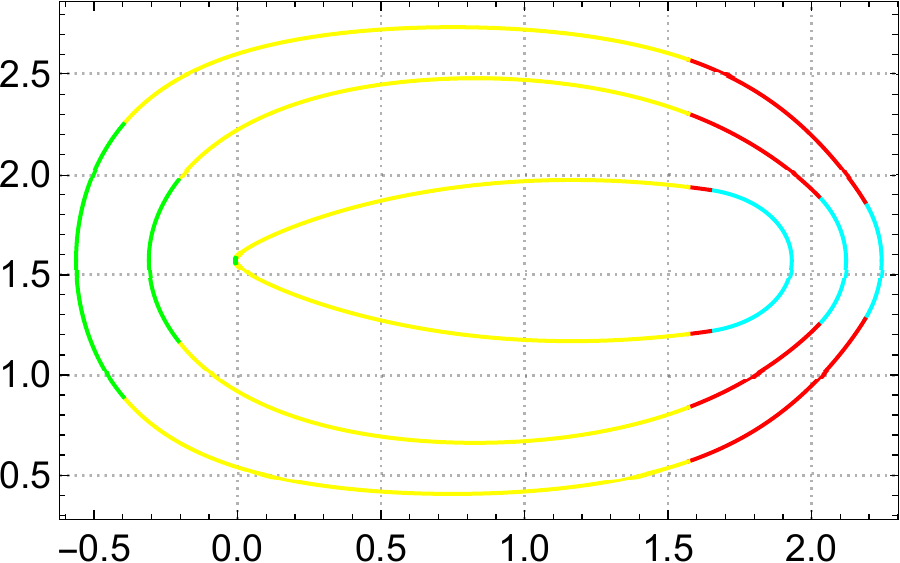}
   \caption{Critical angles for the extreme Kerr-Sen black hole with $a=0.5\,,\theta_*=\pi/3$ in the Carter frame. The vertical axes and horizontal axes stand for $\b_{(i)}$ and $\a_{(i)}$, respectively. $r_*=1.01 r_H, 1.4r_H, 1.8r_H$ from inside to outside. The cyan, red and yellow curves individually stand for the former parts of $\lf(\a_{(i)},\,\b_{(i)}\rt)$ in Eqs. (\ref{eca1}), (\ref{eca2}) and (\ref{ca3}), whose latter parts are represented by the green curves.}
   \label{nonextcone4}
\end{figure}

\begin{figure}[!htbp] 
   \centering
   \includegraphics[width=2.3in]{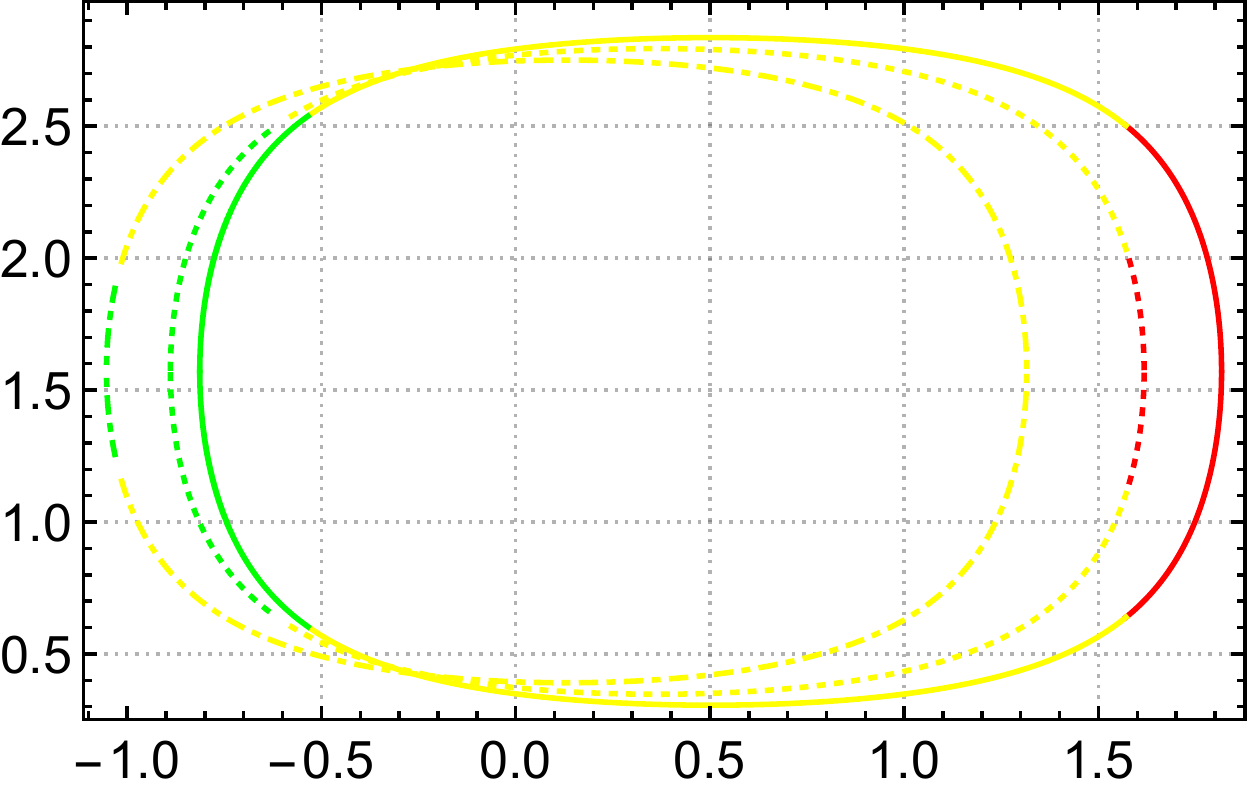}
   \caption{Critical angles for the non-extreme Kerr-Sen black hole with $a=0.4\,,c=0.5\,,r_*=1.5r_H$ in the Carter frame. The vertical axes and horizontal axes stand for $\b_{(i)}$ and $\a_{(i)}$, respectively. $
   \theta=5\pi/12,\,3\pi/12,\,\pi/12$ for the solid, dotted and dot dashed lines, respectively. The red and yellow curves individually stand for the former parts of $\lf(\a_{(i)},\,\b_{(i)}\rt)$ in Eqs. (\ref{neca2}) and (\ref{ca3}), whose latter parts are represented by the green curves.}
   \label{nonextcone3}
\end{figure}

Following similar procedures to the equatorial case, we can calculate the escape cones and the escape probabilities of the photons on the off-equatorial plane and both in the LNRF and Carter frame. Fortunately, we can use the analysis about the radial effective potential $\mathcal{R}$ in Sec. \ref{3}, since it is independent of $\theta_*$. Note that, according to the definitions of the emission angles (\ref{ans1}), (\ref{ans2}), (\ref{ans3}) and (\ref{ans4}), $\theta_*$ enters the calculation as $p^{(t)},\,p^{(\theta)},\,p^{(\phi)}$ are dependent of it in both the LNRF and the Carter frame.  We show representative  escape cones in Figs. \ref{nonextcone4} and \ref{nonextcone3}, from which we can see that the contours of the cones are indeed similar to those for the photons emitted from the source on the equatorial plane in the LNRF.

\begin{figure*}[!htbp] 
   \centering
   \includegraphics[width=2.3in]{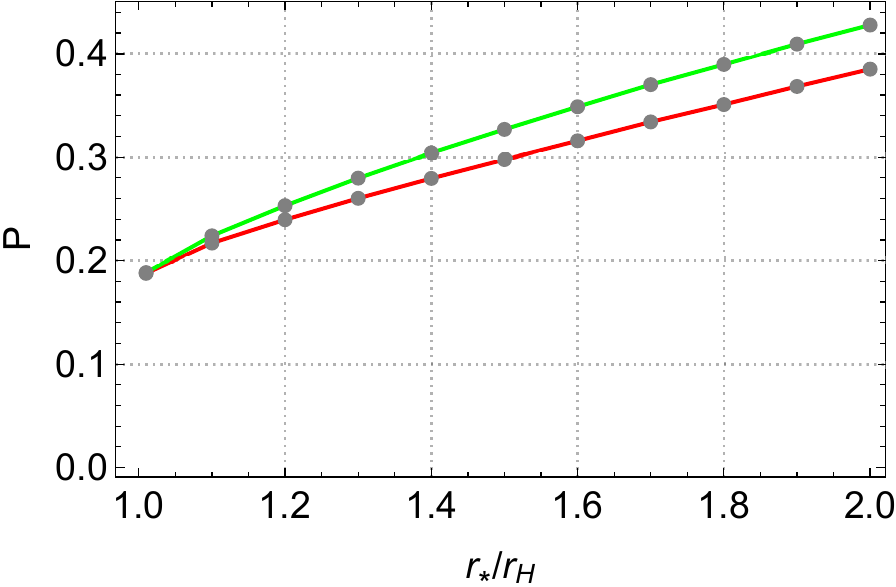}
     \includegraphics[width=2.3in]{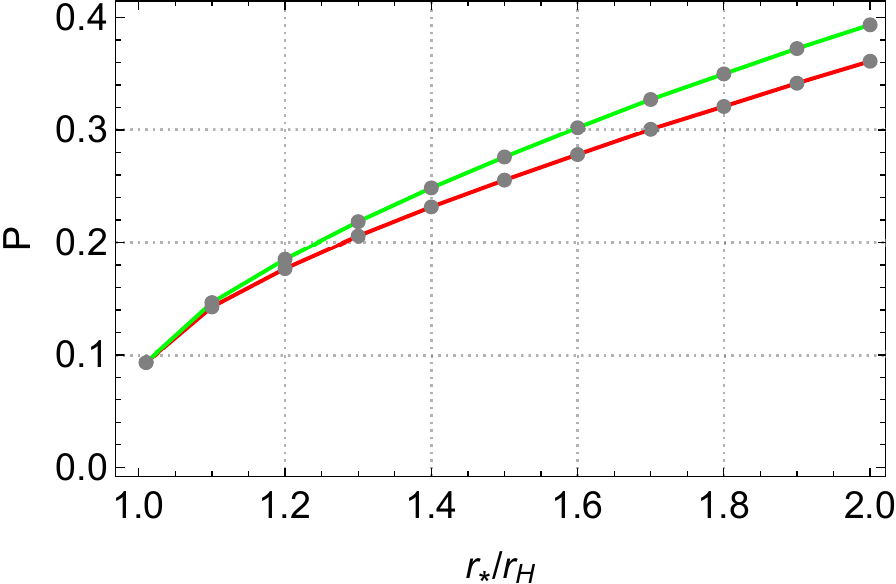}\\
       \includegraphics[width=2.3in]{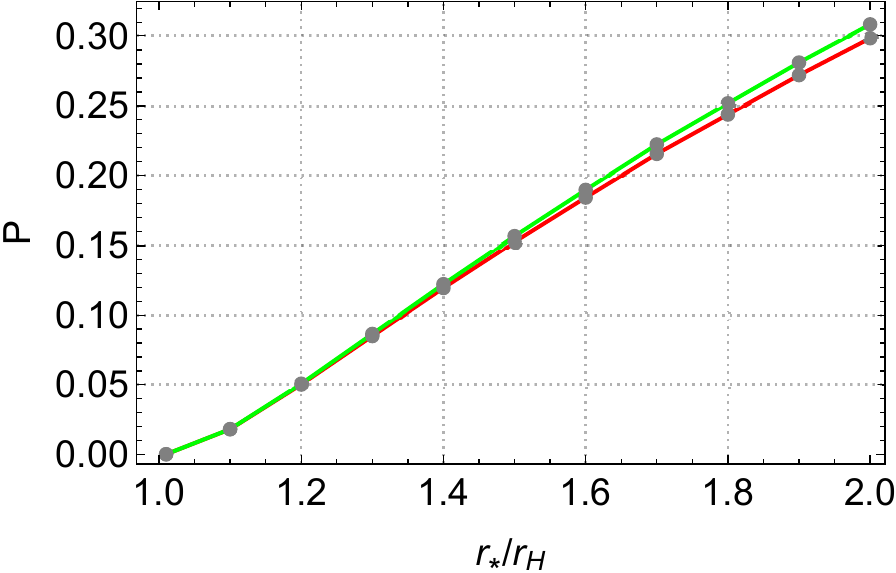}
     \includegraphics[width=2.3in]{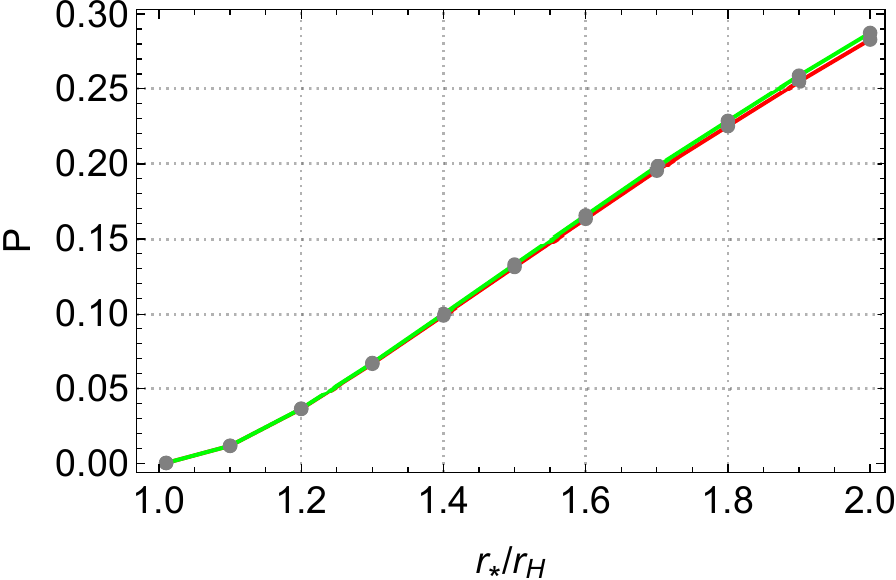}
   \caption{Comparison of the escape probabilities of photons in LNRF (red) and in Carter frame (green) in the background of extreme Kerr-Sen black hole with $a=0.5$. The top two diagrams are for $\theta_*=\pi/2,\,\pi/3$ and the bottom two diagrams are for $\theta_*=\pi/6\,,\pi/9$, respectively.}
   \label{exC1}
\end{figure*}

\begin{figure*}[!htbp] 
   \centering
   \includegraphics[width=2.3in]{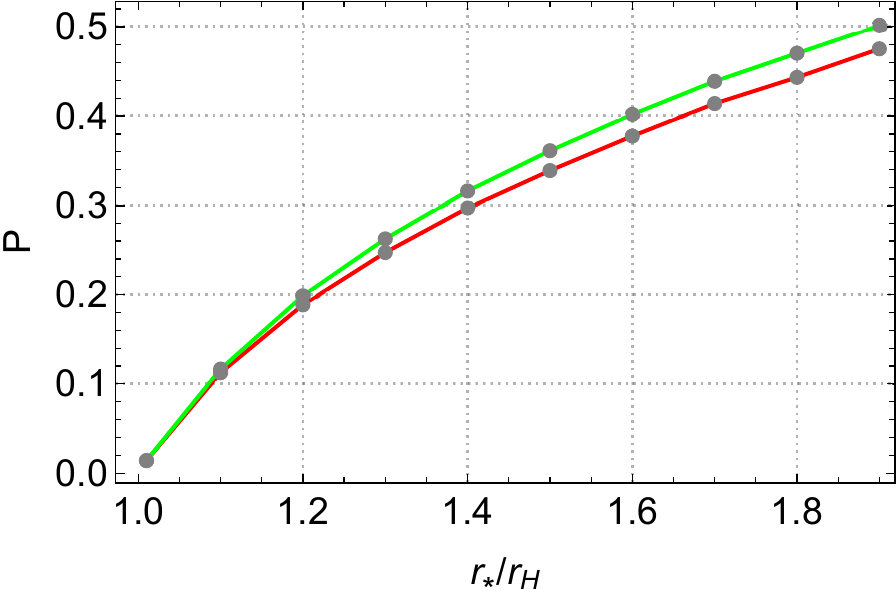}
     \includegraphics[width=2.3in]{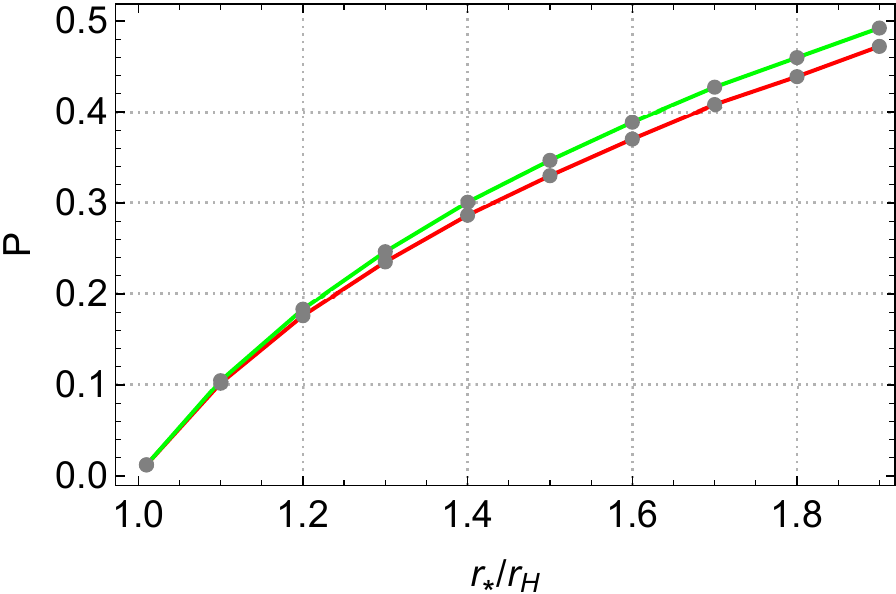}\\
       \includegraphics[width=2.3in]{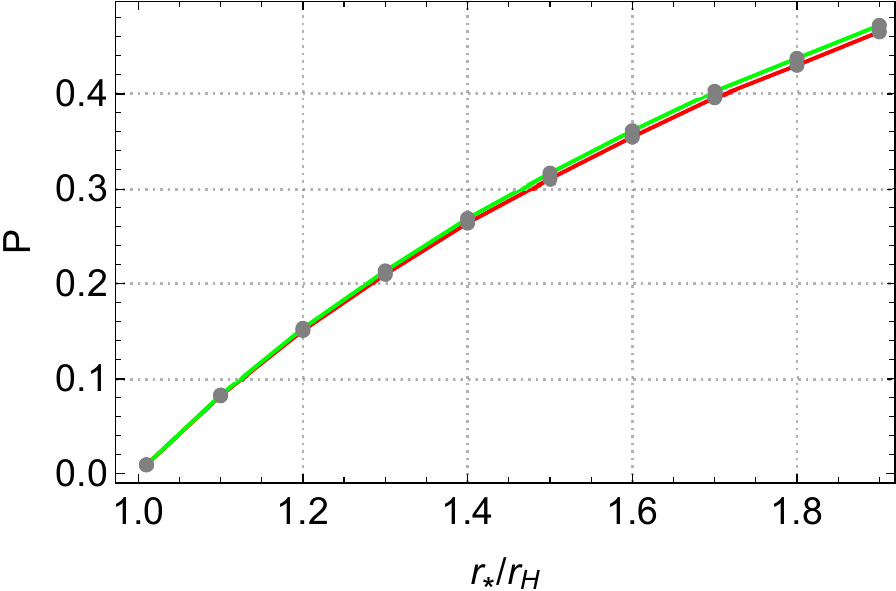}
     \includegraphics[width=2.3in]{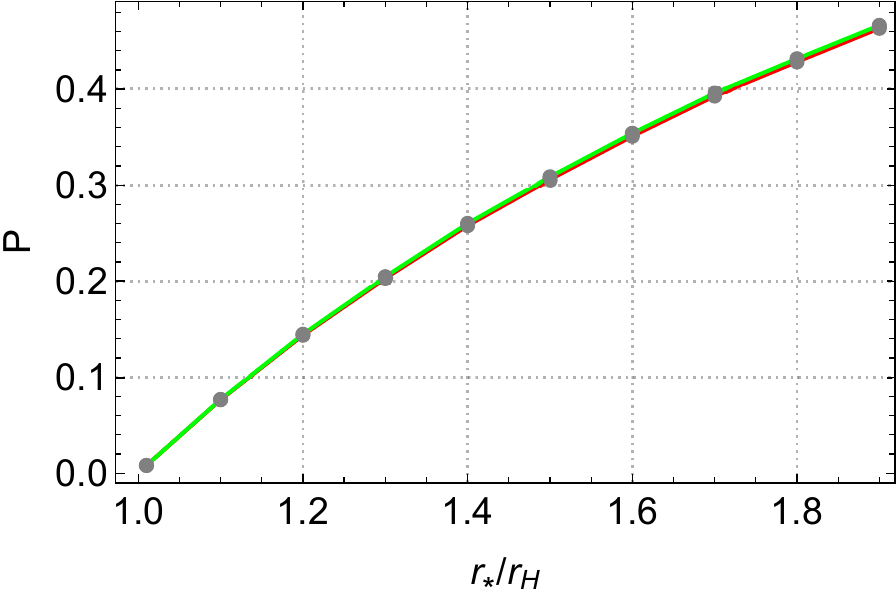}
   \caption{Comparison of the escape probabilities of photons in LNRF (red) and in Carter frame (green) in the background of non-extreme Kerr-Sen black hole with $c=0.5,\,a=0.4$. The top two diagrams are for $\theta_*=\pi/2,\,\pi/3$ and the bottom two diagrams are for $\theta_*=\pi/6\,,\pi/9$, respectively.}
   \label{nexC1}
\end{figure*}

\begin{figure*}[!htbp] 
   \centering
   \includegraphics[width=2.3in]{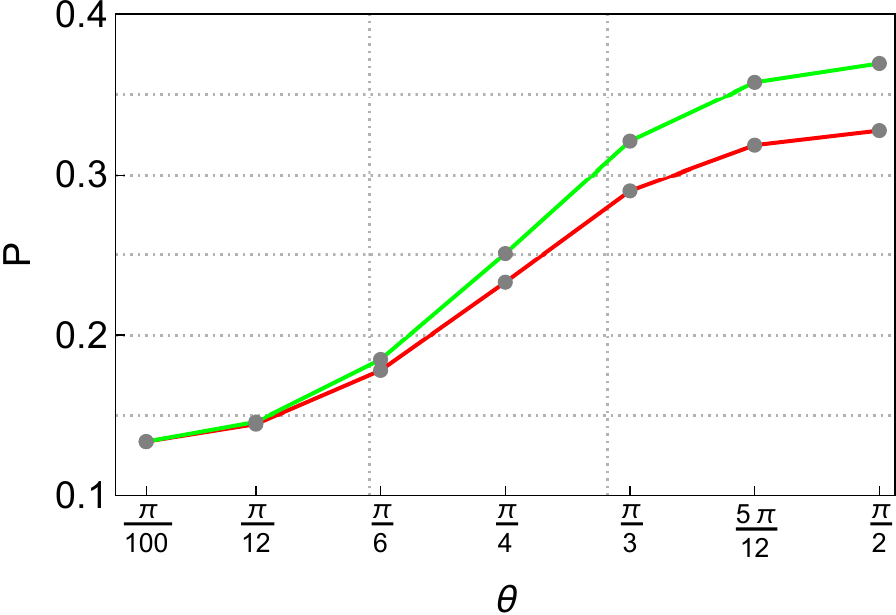}
      \includegraphics[width=2.3in]{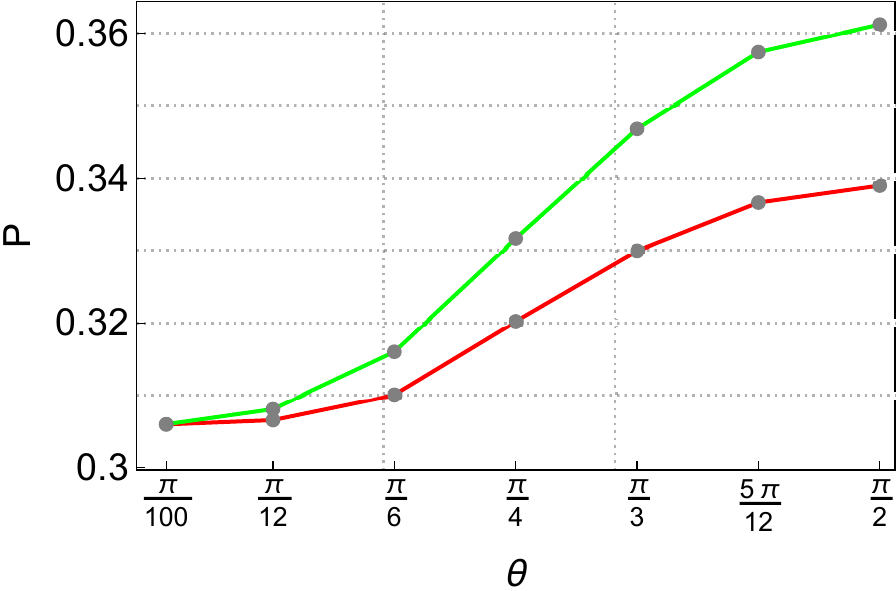}   
   \caption{Comparison of the escape probabilities of photons in LNRF (red) and in Carter frame (green) in the background of  extreme Kerr-Sen black hole (left diagram) with $a=2/3\,,r_*=1.5\,r_H$  and in the background of non-extreme Kerr-Sen black hole (right diagram) with $c=0.5,\,a=0.4\,,r_*=1.5\,r_H$.}
   \label{thetava}
\end{figure*}

To quantitatively study the emission of the photons around the Kerr-Sen black hole, we calculate the escape probabilities of the photons and show them in the Figs. \ref{exC1}, \ref{nexC1} and \ref{thetava}. From Fig. \ref{exC1}, we can see that the escape probability of the photon increases with the radial position of the light source, irrespective of the frame we choose. However, the escape probabilities in the Carter frame are always greater than those in the LNRF. We also note that the escape probability of the photon at the horizon limit of the extreme Kerr-Sen black hole vanishes if the latitude position approaches the pole of the black hole. There is a subtle property we should explain here. For the extreme Kerr-Sen black hole, though the escape probabilities become very small near the horizons for the bottom two cases in Fig. \ref{exC1}, the escape probability of the photon at the horizon limit cannot be zero (besides at the pole, see below). This is evident by analyzing the Fig. \ref{schematic3}a, as the photons can be scattered by the black hole despite being with initially inward velocity.

From Fig. \ref{nexC1}, we can see that the variation trend of the escape probability for the photon with respect to the radial position of the photon source in the non-extreme Kerr-Sen black hole background is similar to the extreme Kerr-Sen black hole case, and the escape probabilities of the photons in the Carter frame are also greater than those in the LNRF. In the horizon limit, the escape probabilities become vanishing both for the latitude position approaching the equator and for the latitude position approaching the pole.

From Fig. \ref{thetava}, we can know that the escape probability of the photon with a constant radial position is maximal on the equatorial plane and it decreases monotonically if the latitude position decreases to $\theta_*=0$ or increases to $\theta_*=\pi$. Likewise, the escape probabilities of the photons in the Carter frame are greater than those in the LNRF. It is noticeable that the difference of the probabilities in the two different frames is maximum on the equatorial plane and when the light source moves to the two poles of the rotating Kerr-Sen black hole, the difference of the probabilities in the two frames becomes smaller and smaller.

We should take care of the cases $\theta_*=0$ and $\theta_*=\pi$, as $p^{\phi}$ vanishes at these two points. Then we only have $\beta$, which can be obtained from
\begin{subequations}
 \begin{align}
\sin\b&=\frac{p^{(r)}}{p^{(t)}}, \label{49a}\\
\cos\b&=-\frac{p^{(\q)}}{p^{(t)}}\label{49b}.
\end{align}
\end{subequations}
$\pi/2-\beta$ at this circumstance means the angle of inclination relative to the radially outward normal vector on the pole of the Kerr-Sen black hole. The escape cone and the escape probability of photon at the pole can be obtained by using the former calculations and taking the pole-limit. By these procedures, we can not only avoid  overestimating the escaping angle \cite{Galtsov:2019bty} but also distinguish the extreme and non-extreme Kerr-Sen black holes cases (Note that in the extreme case $r_1>r_H$ so that $r_1$ can not correspond to the critical  escape angle). As $p^{(r)}$ and $p^{(\theta)}$ are the same in LNRF and Carter frame, we can infer that the escape cone and the escape probability of the photon will also be the same at the pole, which has been shown in Fig. \ref{thetava}, where the blue and the red lines almost coincide with each other near the pole.

With the help of Eqs. (\ref{49a}) and (\ref{49b}), we can calculate the escape angles for the photons at the pole and at the horizon limit. Straightforwardly, we can have 
$$\sin\beta \left(r_*\to r_H,\,\theta_*\to 0\right)=1$$ and
$$\cos\beta \left(r_*\to r_H,\,\theta_*\to 0\right)=0$$
after substituting Eqs. (\ref{yy1}), (\ref{yy2}) and (\ref{ptheta2}) for the LNRF  or  Eqs. (\ref{xx1}), (\ref{xx2}) and (\ref{ptheta}) for the Carter frame into Eqs. (\ref{49a}) and (\ref{49b}). Then we can know that the inclination angle relative to the radially outward normal vector on the pole for the escaping photon should be $0$. It results in the vanishing escape cones and escape probabilities for the photons in both the LNRF and the Carter frames. Together with Figs. \ref{exC1} and \ref{thetava}, we can confirm that the escape probabilities of the photons at the event horizon limit decrease monotonically from its maximum value at $\theta_*=\pi/2$ to zero at the poles ($\theta_*=0,\,\pi$).

\section{Closing remarks}\label{4}
The observations of the black hole shadow, high-energy physics and electromagnetic spectrum accompanied by gravitational waves from the black hole merger to some extent rely on the escape probabilities of the particles from the black hole. We investigated the escape probabilities of the photons from the Kerr-Sen black hole, inspired by the work in Ref. \cite{Ogasawara:2019mir} where the escape probabilities of the photons in the Kerr-Newman spacetime were explored.

Firstly, the light source was set to be at rest in a LNRF on the equatorial plane of the Kerr-Sen black hole. After defining the emission angles, we calculated the escape cones and escape probabilities of the isotropically emitted photons both from extreme and nonextreme Kerr-Sen black holes. As a result, we found that, under the extreme Kerr-Sen black hole background, the escape cones are nonvanishing and the escape probabilities are nonzero in the event horizon limit; in the nonextreme Kerr-Sen black hole background, the escape cones shrink to be null and the escape probabilities become zero in the event horizon limit. Our result makes it clear that the near horizon physics of the extreme Kerr-Sen black hole is more visible than that of the non-extreme Kerr-Sen black hole on the equatorial plane.

Secondly, we further developed the investigation in two aspects: comparing the emission of the photons in the LNRF with the one in the Carter frame and calculating the escape probability of the photon emitted from a source on the off-equatorial plane. We found that the escape probabilities of the photons in the Carter frame are always greater than the ones in the LNRF, except that the light source is at the pole (the probabilities will be equal if the light source locates at the pole).  Note that the observer in the LNRF is also dubbed as a zero angular momentum observer (ZAMO). The ZAMOs' world lines are orthogonal to the hyper-surfaces of the time coordinates. The ZAMO in rest with respect to the LNRF is dragged by the rotation of the black hole in azimuthal direction with an angular velocity $\Omega_{\text{LNRF}}=-g_{t\phi}/g_{\phi\phi}$ relative to a distant static observer. The Carter frame makes it possible to separate the geodesic equations of the particle moving in the rotating Kerr-Sen spacetime. The Carter observers travels around the Kerr-Sen black hole with an angular velocity $a/(r^2+2cr+a^2)$ \cite{znajek1977black} and constant $r$ and $\theta$. Besides, the Carter frame is useful to confirm the algebraic properties of the Kerr-Sen spacetime's curvature tensor \cite{Bini:2017slb}. The ZAMO relates with the observer in the Carter frame by boosts in the azimuthal angular direction which is accompanied by the rotational Killing vector field, so it means that the light sources in these two frames are different on relative azimuthal motion \cite{Bini:2017slb}.   We observed that the escape probabilities of the photons increase with the radial position, irrespective of the orthonormal tetrad we choose and even on the off-equatorial plane. We also showed that the particle on the equatorial plane gains the maximal escape probability and the probability decreases monotonically toward the pole of the Kerr-Sen black hole.  The result shows that different  relative azimuthal motions of the observers correspond to different escape probabilities. The result also indicates that the physical prospects ( such as the high-energy events \cite{Zhang:2020cpu}, as mentioned in the Sec. \ref{introduction}) are more visible to the Carter observer than to the ZAMO.

We evaluated the escape probabilities of photons directly by calculating the area of the warped escape cones, which is different from the way used in Ref. \cite{Ogasawara:2019mir} in which intermediate variables $r_{1}$ and $r_{2}$ where the impact parameters get extreme values were used.

Here let us see how the Kerr-Sen black hole case deviates from the classical general relativity.  We can compare the escape probability of the photon on the equatorial plane of the extreme Kerr-Sen black hole shown in Fig. \ref{epexa} with the one in the extreme Kerr-Newman black hole background shown in Table II of Ref. \cite{Ogasawara:2019mir}. For instance, when $a=0.7$, we have $P=0.221$ for the former  and $P=0.194$ for the latter. The difference is minor. For the non-extreme black hole case, there is not a qualitative difference of escape probabilities between these two kinds of spacetime background cases either.  Due to the similarity of the Kerr-Newman line element and the Kerr-Sen line element, we suspect that our results of the escape cones and escape probabilities  for the photons on the off-equatorial plane of the Kerr-Sen black hole can be applied to the Kerr-Newman case. However, there is still a significant difference. When $0\leqslant a\leqslant 1/2$, the horizon limit of the photon escape probability is nonzero in the extreme Kerr-Sen black hole background but the probability becomes zero in the background of the extreme Kerr-Newman black hole background. The similarities and differences of the escape cones and escape probabilities origin mainly from, we think, the radial effective potential of the particle in the spacetime. The latitude angular effective potential provides subordinate restrict condition.

One perplexing and intriguing thing is that in the horizon limit of the extreme Kerr-Sen black hole, the escape probabilities of the photons approach zero, just like the non-extreme Kerr-Sen black hole case, if the latitude position of the light source is at the pole.  It may be instructive to study the emission of photons from black objects in other modified gravities and the rules of the photon emission may become clearer by comparison.



\section*{Acknowledgements}
Jie Jiang  is supported by the National Natural Science Foundation of China (Grants No. 11775022 and
11873044). Ming Zhang is supported by the Initial Research Foundation of Jiangxi Normal University with Grant No. 12020023.

\bibliography{Notes}

\end{document}